# A generalized GN-model closed-form formula

**Pierluigi Poggiolini**

*Senior Member, IEEE*
*Fellow, Optical Society of America (OSA)*
Politecnico di Torino, Italy
pierluigi.poggiolini@polito.it

**Abstract:** the GN-model of fiber non-linearity has had quite substantial success in modern optical telecommunications network as a design and management tool. A version of it, capable of handling arbitrary WDM combs and link structures in closed form, was proposed in 2014. Here we upgrade that formula, to make it capable of handling frequency-dependent dispersion, frequency-dependent loss and frequency-dependent gain/loss due to Stimulated Raman Scattering (SRS) among channels. This way, more challenging and complex network scenarios, like the ones that are being deployed right now, can be dealt with in real-time, to the great advantage of management, control and optimization of such networks.

## 1. Introduction

The problem of modeling the impact of non-linear effects in an optical fiber has been given substantial attention since the onset of optical fiber communications. Over the last ten years, and manifestly since the introduction of coherent optical systems, substantial results have been achieved. Currently, several NLI (non-linear-interference) models are available, with different features in terms of accuracy vs. complexity, for instance [1]-[7].

One of the most well-known and possibly the most widely used in the industry is the so-called "GN-model". Interestingly, while its name and its current evolved form(s) are relatively recent, the first derivation of an NLI model along the same general lines as the GN-model dates back to 1994 [8]. For a comprehensive introduction to the GN-model, its history, evolution, validation and limitations, please see [9],[10].

One of the most successful variants of the model has been the so-called incoherent GN-model, or iGN model. The model assumes that the NLI produced in each span of an optical transmission systems sums *in power* at the end of the link. All coherent interference at the field level is neglected. While this may seem a crude approximation, given the nature of NLI in modern optical WDM systems, the model accuracy is only modestly impacted. In fact, due to an often-cited error-cancelation effect, in practice no loss of accuracy is typically found. In contrast, the incoherent assumption makes the iGN-model much simpler to deal with analytically.

Based on the iGN model, various closed-form formulas were derived, for instance in [11]. Several of these closed-forms were however derived to explore fundamental limits and as such they typically assumed all-identical spans and/or identical equally-spaced WDM channels.

A closed-form iGN-model version capable of handling *arbitrary* WDM combs and *arbitrary* link structures in closed form, was proposed in [4]. The intention there was to provide the network manufacturers and operators with a flexible tool, capable of handling complex systems in real-time, to the purpose of on-the-fly management, control and optimization. Such formula and its variants are currently in use in actual commercial long-haul high-capacity network equipment from major vendors, such as CISCO Systems.

Interestingly, the incoherent accumulation assumption allows to deal with certain key network *global* characterization and optimization tasks based on *span-local* characterization and optimization. This is the so-called LOGO (Local Optimization – Global Optimization) principle which was first introduced in [12]. This technique, too, is currently in use in commercial equipment.

However, lately, systems have evolved substantially and the need for a more flexible, closed-from iGN formula, capable of handling a wider variety of system scenarios, has become quite urgent. In particular, the current trend towards using challenging fibers, with very low or even zero in-band dispersion, and/or extending the used fiber bandwidth to C+L bands, leads to systems that cannot be dealt with accurately with the previous closed-form of the iGN model.

Specifically, over large bands such as C+L (bordering on 80 nm, or about 10 THz), fiber loss can no longer be considered frequency-independent. The same is true for fiber dispersion, especially in low-dispersion fibers. Finally, again the push to C+L makes it necessary to account for the power-transfer from higher-frequency channels to lower-frequency ones, mediated through SRS (Stimulated Raman Scattering, [13], [14]).

In this paper, we carry out a detailed derivation of a new iGN model formula, accounting for all of the above additional features. This way, more challenging and complex network scenarios, like the ones that are now being deployed, and will be in the near- and mid-term, can be dealt with from the viewpoint of NLI impact assessment, in real-time, to the great advantage of management, control and optimization of such networks.

**Important notice:**

It has come to the author's attention that a similar effort, with similar goals to this paper, had resulted in the posting of [15] on ArXiv. The posting of [15] took place about a month earlier than the first posting of this document and therefore [15] represents the first version of a fully-closed-form NLI formula based on the GN-model. The work in ref. [15] and in this paper have been independently carried out. This has been mutually acknowledged by the respective authors. The final results bear substantial similarities and some differences, whose significance will be investigated in forth-coming documents.

Ref. [15] substantially extends and generalizes [16] and contains some extensive validation of the closed-form formula presented there. It is therefore a quite important contribution as it provides strong confirmation of the general viability of comprehensive GN-model fully closed-form formulas, such as proposed in ref. [15] itself and here, capable of handling arbitrary ultra-broad-band WDM systems.

**This is version 2 of this document.**

The changes in version 2 vs. version 1 are detailed below.

(1): The notice above was added.

(2): References [15] and [16] were added

(3): Acknowledgements were added at the bottom of this paper.

No other change was made to the original posting of version 1.

## 2. Premises

In the first order Regular Perturbation (RP1) paradigm [], NLI noise is produced in each span independently of the NLI produced in any other span. We then add the approximation of *incoherent* NLI accumulation, that is that the NLI produced in each span sums up in *power* at the end of the link. Based on these assumptions, the power spectral density (PSD) of NLI at the end of the link, $G_{\text{NLI}}^{\text{end}}(f)$ is then:

$$G_{\text{NLI}}^{\text{end}}(f) \approx \sum_{n_s=1}^{N_s} G_{\text{NLI}}^{(n_s),\text{end}}(f)$$

*Eq. 1*

where $N_s$ is the number of spans in the link and the "approximately equal" symbol is used here to point out that this is an approximation. However, henceforth we will use an "equal-to" symbol for ease of readability.

The problem of NLI estimation then becomes that of assessing the contribution to NLI at the end of the link, due to the NLI produced in each single span, that is assessing $G_{\text{NLI}}^{(n_s),\text{end}}(f)$ for each $n_s$. Note that $G_{\text{NLI}}^{(n_s),\text{end}}(f)$ represents the NLI noise produced at the $n_s$-th span and *then propagated linearly till the end of the link*. According to the stated approximations, this quantity relates to the NLI PSD produced at the $n_s$-th span and assessed at the end of the same $n_s$-th span, $G_{\text{NLI}}^{(n_s)}(f)$, as follows:

$$G_{\text{NLI}}^{(n_s),\text{end}}(f) = G_{\text{NLI}}^{(n_s)}(f) \cdot \left|\text{H}(f; n_s+1, N_s)\right|^2$$

*Eq. 2*

where $\text{H}(f; n_s, n_s')$ is the linear link transfer function between from the *input* of the $n_s$-th span to the *output* of the $n_s'$-th span, at frequency $f$. Hence, Eq. 1 can be re-written as:

$$G_{\text{NLI}}^{\text{end}}(f) = \sum_{n_s=1}^{N_s} G_{\text{NLI}}^{(n_s)}(f) \cdot \left|\text{H}(f; n_s+1, N_s)\right|^2$$

*Eq. 3*

Note that for the summation term with $n_s = N_s$ the resulting factor has value:

$$\left|\text{H}(f; N_s+1, N_s)\right|^2 = 1$$

This is actually not an artificial assumption. The writing $\left|\text{H}(f; N_s+1, N_s)\right|^2$ means that we are placing the output coincident with the input (the input of a possible $(N_s+1)$-th span would coincide with the output of the $N_s$-th span) and therefore there is no change in the signal.

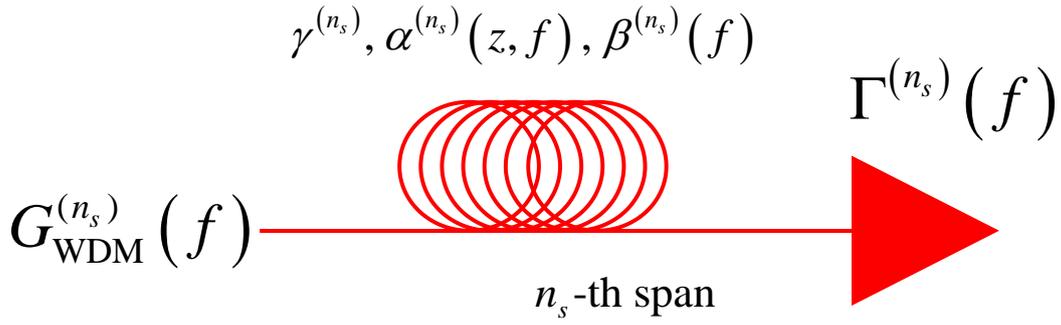

*Figure 1: Pictorial representation of a generic span. A span comprises a single fiber type characterized through its parameters $\gamma^{(n_s)}, \alpha^{(n_s)}(z,f), \beta^{(n_s)}(f)$, a single lumped-elements end section, shown above as a red triangle, which can include: an EDFA, a VOA and a linear filtering element, which overall give rise together to a frequency-dependent power gain/loss $\Gamma^{(n_s)}(f)$. It can also hold a dispersion compensating unit (DCU), although the calculations in the text show that under the assumption made here it is irrelevant to all effects. At the input, there is the WDM overall signal PSD $G_{\mathrm{WDM}}^{(n_s)}(f)$. All quantities are defined in detail in the body of the text.*

We then focus on expressing in general closed-form $\left|\mathrm{H}(f;n_s,n_s')\right|^2$.

What we want to account for in it are the following:

- frequency-dependent fiber dispersion
- frequency-dependent distributed loss/gain
- frequency-dependent lumped loss/gain
- assess the impact of possible lumped dispersion-compensating units (DCUs)

Note that distributed loss/gain may be due to purposeful Raman amplification, or to SRS. In this paper we focus on SRS.

To express $\left|\mathrm{H}(f;n_s,n_s')\right|^2$ we first point out that:

$$\left|\mathrm{H}(f;n_s,n_s')\right|^2 = \prod_{n=n_s}^{n_s'} \left|\mathrm{H}(f;n,n)\right|^2$$

*Eq. 4*

So, we can concentrate on the simpler problem of expressing $\left|\mathrm{H}(f;n,n)\right|^2$, which is the square modulus of the linear transfer function of the $n$-th span. We have:

$$\mathrm{H}(f;n,n) = e^{-j\beta_{\mathrm{DCU}}^{(n)}(f)} e^{\int_0^{L_s^{(n)}} \kappa^{(n)}(f,z)dz} \sqrt{\Gamma^{(n)}(f)}$$

*Eq. 5*

where $\kappa^{(n)}(f,z)$ is a generalized propagation constant defined as:

$$\kappa^{(n)}(f,z) = -j\beta^{(n)}(f,z) - \alpha^{(n)}(f,z)$$

where: $\beta^{(n)}(f,z)$ is the propagation constant (sometime called "wavenumber") for the $n$-th span, at frequency $f$ and location $z$ in the span, in units 1/km; $\alpha^{(n)}(f,z)$ is the fiber loss coefficient for the $n$-th span, at frequency $f$ and location $z$ in the span, in units 1/km; $\beta_{\text{DCU}}^{(n)}(f)$ is a lumped dispersion-compensating element placed at the end of the $n$-th fiber span.

Note that in Eq. 5 it is assumed that the spatial variable $z$ is re-initialized to zero at the start of each span, that is, it is local to the span. It runs from 0 to the $n$-th span length $L_s^{(n)}$. Also, distributed gain is incorporated within $\alpha^{(n)}(f,z)$ for practicality. This means that there may be stretches of fiber whose $\alpha^{(n)}(f,z) < 0$. Finally, $\Gamma^{(n)}(f)$ is frequency-dependent lumped gain/loss for the $n$-th span, such as EDFA, VOA or an optical filter, *placed all at the end of the span*. Note that any lumped gain or loss preceding the $n$-th span fiber must be assigned to the span $(n-1)$.

We now make some assumptions on $\kappa^{(n)}(f,z)$. Firstly, we assume that:

$$\beta^{(n)}(f,z) = \beta^{(n)}(f)$$

i.e., dispersion is assumed to be $z$-independent. This assumption leads to no practical loss of generality, since spans are typically made up of a single type of fiber. If a span is made up of two types of fiber, then it can always be formally split up into two spans, with unit lumped amplification between them. As a result:

$$\kappa^{(n)}(f,z) = -j\beta^{(n)}(f) - \alpha^{(n)}(f,z)$$

Eq. 6

Then we can write:

$$\mathrm{H}(f;n,n) = \sqrt{\Gamma^{(n)}(f)} e^{-j\beta_{\text{DCU}}^{(n)}(f)} e^{\int_0^{L_s^{(n)}} \kappa^{(n)}(f,z) dz} = \sqrt{\Gamma^{(n)}(f)} e^{-j\beta_{\text{DCU}}^{(n)}(f)} e^{-j\beta^{(n)}(f)L_s^{(n)}} e^{-\int_0^{L_s^{(n)}} \alpha^{(n)}(f,z) dz}$$

Eq. 7

We can now take the absolute value squared:

$$\left|\mathrm{H}(f;n,n)\right|^2 = \Gamma^{(n)}(f) e^{-2\int_0^{L_s^{(n)}} \alpha^{(n)}(f,z) dz}$$

Eq. 8

so that Eq. 4 now reads:

$$\left|\mathrm{H}(f;n_s,n_s')\right|^2 = \prod_{n=n_s}^{n_s'} \Gamma^{(n)}(f) e^{-2\int_0^{L_s^{(n)}} \alpha^{(n)}(f,z) dz}$$

Eq. 9

Note the interesting aspect that both distributed fiber dispersion and lumped dispersion compensation disappear from Eq. 9 and Eq. 10.

Eq. 3 then becomes:

$$G_{\text{NLI}}^{\text{end}}(f) = \sum_{n_s=1}^{N_s} G_{\text{NLI}}^{(n_s)}(f) \cdot \prod_{n=n_s+1}^{N_s} \Gamma^{(n)}(f) e^{-2\int_0^{L_s^{(n)}} \alpha^{(n)}(f,z)dz}$$

*Eq. 10*

This equation embodies our fundamental premises, including the incoherent NLI accumulation assumption.

### 3. The NLI produced in the generic *n*-th span

The next step is calculating the NLI produced in the generic $n$-th span. As a notation choice, the index used to identify spans will henceforth be written $n_s$, where the subscript is a reminder for "span". The reason is to not confuse this index with other indices that will be used in the derivation.

According to the assumptions and derivations of the incoherent GN model (iGN model), we have for the generic $n_s$-th span NLI PSD $G_{\text{NLI}}^{(n_s)}(f)$ the following general formula [4],[9]:

$$G_{\text{NLI}}^{(n_s)}(f) = \frac{16}{27} \int_{-\infty}^{\infty}\int_{-\infty}^{\infty} G_{\text{WDM}}^{(n_s)}(f_1) G_{\text{WDM}}^{(n_s)}(f_2) G_{\text{WDM}}^{(n_s)}(f_1+f_2-f)$$

$$\left| H(f;n_s,n_s) \int_0^{L_s^{(n_s)}} \gamma^{(n_s)} e^{\int_0^z \left[-\kappa^{(n_s)}(f,z')+\kappa^{(n_s)}(f_2,z')+\kappa^{(n_s)*}(f_1+f_2-f,z')+\kappa^{(n_s)}(f_1,z')\right]dz'} dz \right|^2 df_1 df_2$$

The quantity $G_{\text{WDM}}^{(n_s)}(f)$ represents the PSD of the WDM signal at the input of the $n_s$-th span. An example is shown in Fig. 2.

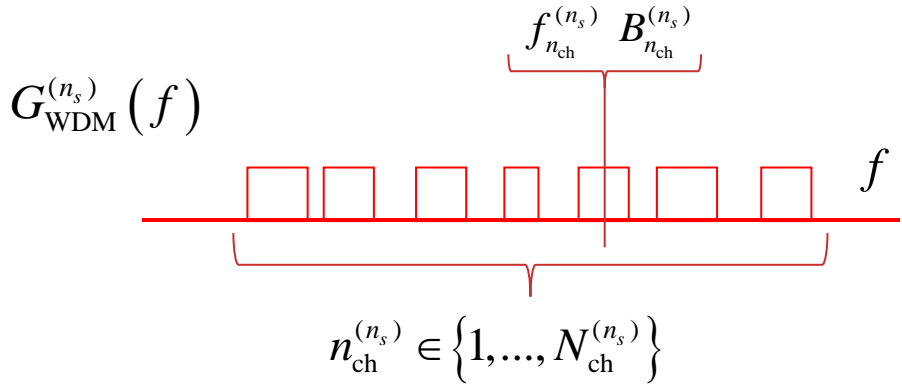

*Figure 2: Pictorial representation of an example of a WDM spectrum $G_{\text{WDM}}^{(n_s)}(f)$. Symbols are explained in the text.*

The symbol $\gamma^{(n_s)}$ represents the $n_s$-th span fiber non-linearity coefficient, typically expressed as 1/(W km). Recalling Eq. 7 and assuming that $\gamma^{(n_s)}$ is $z$-independent and independent of frequency:

$$G_{\text{NLI}}^{(n_s)}(f) = \frac{16}{27}\left[\gamma^{(n_s)}\right]^2 \Gamma^{(n_s)}(f) \int_{-\infty}^{\infty}\int_{-\infty}^{\infty} G_{\text{WDM}}^{(n_s)}(f_1) G_{\text{WDM}}^{(n_s)}(f_2) G_{\text{WDM}}^{(n_s)}(f_1+f_2-f)$$

$$\left| e^{-j\beta_{\text{DCU}}^{(n_s)}(f)} e^{-j\beta^{(n_s)}(f)L_s^{(n)}} e^{-\int_0^{L_s^{(n_s)}}\alpha^{(n_s)}(f,z)dz} L_s^{(n_s)} \int_0^z e^{\int_0^z\left[-\kappa^{(n_s)}(f,z')+\kappa^{(n_s)}(f_2,z')+\kappa^{(n_s)*}(f_1+f_2-f,z')+\kappa^{(n_s)}(f_1,z')\right]dz'} dz \right|^2 df_1 df_2$$

The dispersive terms in absolute value squared factorize to 1 so that we can simplify:

$$G_{\text{NLI}}^{(n_s)}(f) = \frac{16}{27}\left[\gamma^{(n_s)}\right]^2 \Gamma^{(n_s)}(f) \int_{-\infty}^{\infty}\int_{-\infty}^{\infty} G_{\text{WDM}}^{(n_s)}(f_1) G_{\text{WDM}}^{(n_s)}(f_2) G_{\text{WDM}}^{(n_s)}(f_1+f_2-f)$$

$$\left| e^{-\int_0^{L_s^{(n_s)}}\alpha^{(n_s)}(f,z)dz} L_s^{(n_s)} \int_0^z e^{\int_0^z\left[-\kappa^{(n_s)}(f,z')+\kappa^{(n_s)}(f_2,z')+\kappa^{(n_s)*}(f_1+f_2-f,z')+\kappa^{(n_s)}(f_1,z')\right]dz'} dz \right|^2 df_1 df_2$$

*Eq. 11*

This means that under the iGN model assumptions, lumped dispersion has no effect on NLI generation. Henceforth, therefore, we will completely disregard DCUs.

Then, from Eq. 6 :

$$\kappa^{(n_s)}(f_1,z) + \kappa^{(n_s)}(f_2,z) + \kappa^{(n_s)*}(f_1+f_2-f,z) - \kappa^{(n_s)}(f,z) =$$
$$= -j\beta^{(n_s)}(f_1) - \alpha^{(n_s)}(f_1,z) - j\beta^{(n_s)}(f_2) - \alpha^{(n_s)}(f_2,z)$$
$$+ j\beta^{(n_s)}(f) + \alpha^{(n_s)}(f,z) + j\beta^{(n_s)}(f_1+f_2-f) - \alpha^{(n_s)}(f_1+f_2-f,z)$$
$$= -j\left[\beta^{(n_s)}(f_1) + \beta^{(n_s)}(f_2) - \beta^{(n_s)}(f) - \beta^{(n_s)}(f_1+f_2-f)\right]$$
$$-\left[\alpha^{(n_s)}(f_1,z) + \alpha^{(n_s)}(f_2,z) - \alpha^{(n_s)}(f,z) + \alpha^{(n_s)}(f_1+f_2-f,z)\right]$$

For ease of notation, we define:

$$\Delta\beta^{(n_s)}(f_1,f_2,f) = \beta^{(n_s)}(f_1) + \beta^{(n_s)}(f_2) - \beta^{(n_s)}(f) - \beta^{(n_s)}(f_1+f_2-f)$$
$$\Delta\alpha^{(n_s)}(f_1,f_2,f,z) = \alpha^{(n_s)}(f_1,z) + \alpha^{(n_s)}(f_2,z) - \alpha^{(n_s)}(f,z) + \alpha^{(n_s)}(f_1+f_2-f,z)$$

*Eq. 12*

so that:

$$\kappa^{(n_s)}(f_1,z) + \kappa^{(n_s)}(f_2,z) + \kappa^{(n_s)*}(f_1+f_2-f,z) - \kappa^{(n_s)}(f,z) =$$
$$= -j\Delta\beta^{(n_s)}(f_1,f_2,f) - \Delta\alpha^{(n_s)}(f_1,f_2,f,z)$$

*Eq. 13*

This way, we can rewrite Eq. 11 as:

$$G_{\text{NLI}}^{(n_s)}(f) = \frac{16}{27}\left[\gamma^{(n_s)}\right]^2 \Gamma^{(n_s)}(f) \int_{-\infty}^{\infty}\int_{-\infty}^{\infty} G_{\text{WDM}}^{(n_s)}(f_1) G_{\text{WDM}}^{(n_s)}(f_2) G_{\text{WDM}}^{(n_s)}(f_1+f_2-f)$$

$$\left| e^{-\int_0^{L_s^{(n_s)}} \alpha^{(n_s)}(f,z)dz} \int_0^{L_s^{(n_s)}} e^{\int_0^z \left[-j\Delta\beta^{(n_s)}(f_1,f_2,f) - \Delta\alpha^{(n_s)}(f_1,f_2,f,z')\right]dz'} dz \right|^2 df_1 df_2$$

*Eq. 14*

Since $\Delta\beta$ does not depend on $z$, we have:

$$G_{\text{NLI}}^{(n_s)}(f) = \frac{16}{27}\left[\gamma^{(n_s)}\right]^2 \Gamma^{(n_s)}(f) \int_{-\infty}^{\infty}\int_{-\infty}^{\infty} G_{\text{WDM}}^{(n_s)}(f_1) G_{\text{WDM}}^{(n_s)}(f_2) G_{\text{WDM}}^{(n_s)}(f_1+f_2-f)$$

$$\left| e^{-\int_0^{L_s^{(n_s)}} \alpha^{(n_s)}(f,z)dz} \int_0^{L_s^{(n_s)}} e^{-j\Delta\beta^{(n_s)}(f_1,f_2,f)z} e^{-\int_0^z \Delta\alpha^{(n_s)}(f_1,f_2,f,z')dz'} dz \right|^2 df_1 df_2$$

*Eq. 15*

We can now proceed to make specific assumptions on the form of $\beta^{(n_s)}$ and $\alpha^{(n_s)}$.

We first assume that dispersion is expressed through its third-order series expansion as:
$$\beta^{(n_s)}(f) =$$
$$\beta_0^{(n_s)} + 2\pi\beta_1^{(n_s)}\left(f - f_c^{(n_s)}\right) + 4\pi^2\beta_2^{(n_s)}\left(f - f_c^{(n_s)}\right)^2 \frac{1}{2!} + 8\pi^3\beta_3^{(n_s)}\left(f - f_c^{(n_s)}\right)^3 \frac{1}{3!} + O\left(f - f_c^{(n_s)}\right)^4$$

*Eq. 16*

where $f_c^{(n_s)}$ is the arbitrary frequency where the expansion is taken in the $n_s$-th span. Substituting Eq. 16 into Eq. 12 we get the remarkably simple expression:

$$\Delta\beta^{(n_s)}(f_1,f_2,f) = -4(f-f_1)(f-f_2)\pi^2\left(\beta_2^{(n_s)} + \pi\beta_3^{(n_s)}\left[f_1+f_2-2f_c^{(n_s)}\right]\right)$$

*Eq. 17*

We then make a key assumption. We assume that loss can be expressed as the sum of a $z$-independent and a $z$-dependent term. The latter decays exponentially. We write:

$$\alpha^{(n_s)}(f,z) = \alpha_0^{(n_s)}(f) + \alpha_1^{(n_s)}(f)\exp\left(-\sigma^{(n_s)}(f)\cdot z\right)$$

We will come back to this hypothesis later, to better justify it.

Substituting into Eq. 12 we get:

$$\Delta\alpha^{(n_s)}(f_1,f_2,f,z) = \alpha^{(n_s)}(f_1,z) + \alpha^{(n_s)}(f_2,z) - \alpha^{(n_s)}(f,z) + \alpha^{(n_s)}(f_1+f_2-f,z)$$
$$= \alpha_0^{(n_s)}(f_1) + \alpha_1^{(n_s)}(f_1)\exp(-\sigma^{(n_s)}(f_1)\cdot z) +$$
$$\alpha_0^{(n_s)}(f_2) + \alpha_1^{(n_s)}(f_2)\exp(-\sigma^{(n_s)}(f_2)\cdot z) -$$
$$\alpha_0^{(n_s)}(f) + \alpha_1^{(n_s)}(f)\exp(-\sigma^{(n_s)}(f)\cdot z) +$$
$$\alpha_0^{(n_s)}(f_1+f_2-f) + \alpha_1^{(n_s)}(f_1+f_2-f)\exp(-\sigma^{(n_s)}(f_1+f_2-f)\cdot z)$$

Eq. 18

In order to proceed further it is necessary to make some remarks on the frequency integrals involved and take some approximations.

We re-write Eq. 15 as:

$$G_{\text{NLI}}^{(n_s)}(f) = \frac{16}{27}\left[\gamma^{(n_s)}\right]^2 \Gamma^{(n_s)}(f) \int_{-\infty}^{\infty}\int_{-\infty}^{\infty} G_{\text{WDM}}^{(n_s)}(f_1) G_{\text{WDM}}^{(n_s)}(f_2) G_{\text{WDM}}^{(n_s)}(f_1+f_2-f) |\rho(f_1,f_2,f)|^2 df_1 df_2$$

Eq. 19

where:

$$|\rho(f_1,f_2,f)|^2 = \left| e^{-\int_0^{L_s^{(n_s)}} \alpha^{(n_s)}(f,z)dz} \int_0^{L_s^{(n_s)}} e^{-j\Delta\beta^{(n_s)}(f_1,f_2,f)z} e^{-\int_0^{z}\Delta\alpha^{(n_s)}(f_1,f_2,f,z')dz'} dz \right|^2$$

Eq. 20

We look at the structure of the WDM signal $G_{\text{WDM}}^{(n_s)}(f_1)$, with reference to Fig. 2. We assume that in each span there can be a different set of WDM channels. However, the channel under test (CUT) is present in each span, at the same frequency. All other channels can change, in number, frequency and bandwidth.

Specifically, the number of WDM channels in the $n_s$-th span is $N_{\text{ch}}^{(n_s)}$. The set of center frequencies and set of channel bandwidths in the $n_s$-th span are:

$$\{f_{n_{\text{ch}}}^{(n_s)}\}_{n_{\text{ch}}=1}^{N_{\text{ch}}^{(n_s)}} \quad , \quad \{B_{n_{\text{ch}}}^{(n_s)}\}_{n_{\text{ch}}=1}^{N_{\text{ch}}^{(n_s)}} \quad .$$

We explicitly write the CUT channel index in the $n_s$-th span as $n_{\text{CUT}}^{(n_s)}$. Of course, $n_{\text{CUT}}^{(n_s)}$ is one of the channel indices of that span, that is $n_{\text{CUT}}^{(n_s)} \in \{1,\ldots,N_{\text{ch}}^{(n_s)}\}$. The frequency and bandwidth of the CUT in the $n_s$-th span should then be written $f_{n_{\text{CUT}}^{(n_s)}}^{(n_s)}$, $B_{n_{\text{CUT}}^{(n_s)}}^{(n_s)}$. However, the frequency and bandwidth of the CUT stay the same throughout the link, so we can write them for ease of notation as fixed constants:

$$f_{n_{\text{CUT}}^{(n_s)}}^{(n_s)} = f_{\text{CUT}} \quad , \quad \forall n_s$$

$$B_{n_{\text{CUT}}^{(n_s)}}^{(n_s)} = B_{\text{CUT}} \quad , \quad \forall n_s$$

Henceforth, we will identify them simply as: $f_{\text{CUT}}$, $B_{\text{CUT}}$, since they do not change span-by-span.

We then approximate the channel spectra with rectangles, whose bandwidth is $B_{n_{\text{ch}}}^{(n_s)}$. We have for the WDM comb at the input of the $n_s$-th span (see Fig. 2):

$$G_{\text{WDM}}^{(n_s)}(f) = \sum_{n_{\text{ch}}=1}^{N_{\text{ch}}^{(n_s)}} G_{\text{WDM},n_{\text{ch}}}^{(n_s)} \cdot \Pi_{B_{n_{\text{ch}}}^{(n_s)}}\left(f - f_{n_{\text{ch}}}^{(n_s)}\right)$$

*Eq. 21*

Introducing Eq. 21 into Eq. 19 we get:

$$G_{\text{NLI}}^{(n_s)}(f) = \frac{16}{27}\left[\gamma^{(n_s)}\right]^2 \Gamma^{(n_s)}(f) \sum_{k_{\text{ch}}=1}^{N_{\text{ch}}^{(n_s)}} G_{\text{WDM},k_{\text{ch}}}^{(n_s)} \sum_{m_{\text{ch}}=1}^{N_{\text{ch}}^{(n_s)}} G_{\text{WDM},m_{\text{ch}}}^{(n_s)} \sum_{n_{\text{ch}}=1}^{N_{\text{ch}}^{(n_s)}} G_{\text{WDM},n_{\text{ch}}}^{(n_s)}$$

$$\int_{-\infty}^{\infty}\int_{-\infty}^{\infty} \Pi_{B_{k_{\text{ch}}}^{(n_s)}}\left(f_1 - f_{k_{\text{ch}}}^{(n_s)}\right) \Pi_{B_{m_{\text{ch}}}^{(n_s)}}\left(f_2 - f_{m_{\text{ch}}}^{(n_s)}\right) \Pi_{B_{n_{\text{ch}}}^{(n_s)}}\left(f_1 + f_2 - f - f_{n_{\text{ch}}}^{(n_s)}\right) \left|\rho(f_1, f_2, f)\right|^2 df_1 df_2$$

We then concentrate on calculating $G_{\text{NLI}}^{(n_s)}(f)$ related to a *single channel*, specifically the CUT. Note that the CUT can be any of the WDM comb channels, so there is actually no loss of generality in this assumption. However, formally, we limit our interest to finding *the PSD of NLI at the center of the CUT*, that is $G_{\text{NLI}}^{(n_s)}(f_{\text{CUT}})$.

Then:

$$G_{\text{NLI}}^{(n_s)}(f_{\text{CUT}}) = \frac{16}{27}\left[\gamma^{(n_s)}\right]^2 \Gamma^{(n_s)}(f_{\text{CUT}}) \sum_{k_{\text{ch}}=1}^{N_{\text{ch}}^{(n_s)}} G_{\text{WDM},k_{\text{ch}}}^{(n_s)} \sum_{m_{\text{ch}}=1}^{N_{\text{ch}}^{(n_s)}} G_{\text{WDM},m_{\text{ch}}}^{(n_s)} \sum_{n_{\text{ch}}=1}^{N_{\text{ch}}^{(n_s)}} G_{\text{WDM},n_{\text{ch}}}^{(n_s)}$$

$$\int_{-\infty}^{\infty}\int_{-\infty}^{\infty} \Pi_{B_{k_{\text{ch}}}^{(n_s)}}\left(f_1 - f_{k_{\text{ch}}}^{(n_s)}\right) \Pi_{B_{m_{\text{ch}}}^{(n_s)}}\left(f_2 - f_{m_{\text{ch}}}^{(n_s)}\right) \Pi_{B_{n_{\text{ch}}}^{(n_s)}}\left(f_1 + f_2 - f_{\text{CUT}} - f_{n_{\text{ch}}}^{(n_s)}\right) \left|\rho(f_1, f_2, f_{\text{CUT}})\right|^2 df_1 df_2$$

*Eq. 22*

Each of the terms of the triple summation then creates a specific integration domain in $f_1$ and $f_2$. Specifically, considering only one of the summation terms, the integration domain for $f_1$ and $f_2$ results from the intersection of three conditions:

$f_1 \in \left[f_{k_{\text{ch}}}^{(n_s)} - B_{k_{\text{ch}}}^{(n_s)}/2, f_{k_{\text{ch}}}^{(n_s)} + B_{k_{\text{ch}}}^{(n_s)}/2\right]$ due to $\Pi_{B_{k_{\text{ch}}}^{(n_s)}}\left(f_1 - f_{k_{\text{ch}}}^{(n_s)}\right)$

$f_2 \in \left[f_{m_{\text{ch}}}^{(n_s)} - B_{m_{\text{ch}}}^{(n_s)}/2, f_{m_{\text{ch}}}^{(n_s)} + B_{m_{\text{ch}}}^{(n_s)}/2\right]$ due to $\Pi_{B_{m_{\text{ch}}}^{(n_s)}}\left(f_2 - f_{m_{\text{ch}}}^{(n_s)}\right)$

$f_1 + f_2 \in \left[f_{\text{CUT}} + f_{n_{\text{ch}}}^{(n_s)} - B_{n_{\text{ch}}}^{(n_s)}/2, f_{\text{CUT}} + f_{n_{\text{ch}}}^{(n_s)} + B_{n_{\text{ch}}}^{(n_s)}/2\right]$ due to $\Pi_{B_{n_{\text{ch}}}^{(n_s)}}\left(f_1 + f_2 - f_{\text{CUT}} - f_{n_{\text{ch}}}^{(n_s)}\right)$

The resulting integration domain is made up of distinct sub-domains. Looking for instance at an example of a 7-channel WDM comb, made up of equally spaced, identical-bandwidth channel, where the CUT is the center channel of 7, the resulting diagram over the $f_1$ and $f_2$ plane would be as follows:

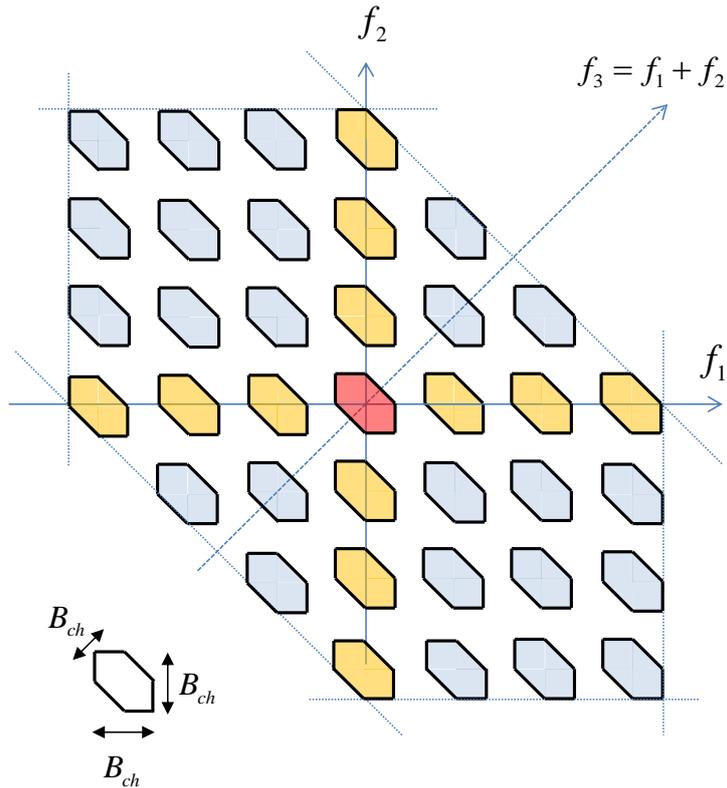

*Figure 3: integration domain "islands" of Eq. 22, for a 7-channel WDM comb of equally spaced and same-bandwidth channels. The CUT is the center channel. In yellow the XCI islands, in red the SCI island.*

The integration domain is made up of several "islands", as shown in Fig.3. Note that if the channel spacing goes below a certain threshold, then smaller triangular sub-islands appear as well [11]. We will in the following neglect the smaller triangular islands. The islands that typically contribute the most to NLI are those along the two axes going through the center frequency of the CUT, i.e., those marked in yellow (the so-called "XCI" or "cross-channel interference islands") and the "SCI" (or "self-channel interference") island at the center, in red (for terminology and explanation, see [11]).

We focus on the red and yellow islands for now. Fig. 4 below serves as a reference for the more general case than Fig.3, where channels are non-equally-spaced and have different bandwidths.

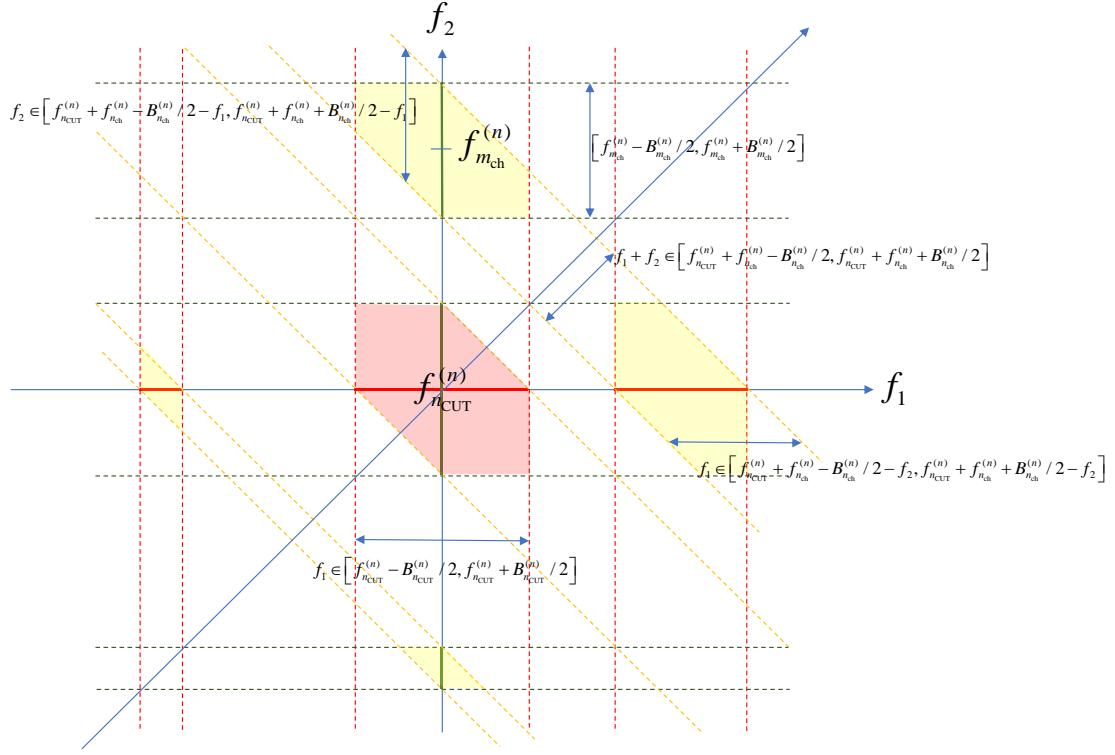

*Figure 4: XCI and SCI integration domain "islands" of Eq. 22, for a 3-channel WDM comb of unequally spaced and different-bandwidth channels. The CUT is the center channel. In yellow the XCI islands, in red the SCI island.*

We first focus on the yellow island along the $f_2$ axis centered at $f_{m_{ch}}^{(n_s)}$. This island is found by limiting integration over $f_1$ due to the CUT spectrum to:

$$f_1 \in \left[ f_{CUT} - B_{CUT}/2, f_{CUT} + B_{CUT}/2 \right]$$

Integration over $f_2$ is then limited by the right side-channel to the CUT, located at $f_{m_{ch}}^{(n_s)}$:

$$f_2 \in \left[ f_{m_{ch}}^{(n_s)} - B_{m_{ch}}^{(n_s)}/2, f_{m_{ch}}^{(n_s)} + B_{m_{ch}}^{(n_s)}/2 \right]$$

There is the third limitation that can be written in three equivalent ways

$$f_2 \in \left[ f_{CUT} + f_{n_{ch}}^{(n)} - B_{n_{ch}}^{(n_s)}/2 - f_1, f_{CUT} + f_{n_{ch}}^{(n_s)} + B_{n_{ch}}^{(n_s)}/2 - f_1 \right]$$

$$f_1 + f_2 \in \left[ f_{CUT} + f_{n_{ch}}^{(n)} - B_{n_{ch}}^{(n_s)}/2, f_{CUT} + f_{n_{ch}}^{(n_s)} + B_{n_{ch}}^{(n)}/2 \right]$$

$$f_1 \in \left[ f_{CUT} + f_{n_{ch}}^{(n)} - B_{n_{ch}}^{(n_s)}/2 - f_2, f_{CUT} + f_{n_{ch}}^{(n_s)} + B_{n_{ch}}^{(n_s)}/2 - f_2 \right]$$

As it is shown in Fig.4, this third limitation causes two things:

- Makes the islands depart from a rectangular shape
- Locks together the $n_{ch}$ and $m_{ch}$ indices or, alternatively, the $n_{ch}$ and $k_{ch}$ so that $n_{ch} = m_{ch}$ or $n_{ch} = k_{ch}$.

Regarding the first bullet, we neglect its effect. If we first assume $n_{ch} = m_{ch}$, then integration will take place over *rectangular* islands defined by:

$$f_1 \in [f_{CUT} - B_{CUT}/2, f_{CUT} + B_{CUT}/2]$$
$$f_2 \in [f_{m_{ch}}^{(n_s)} - B_{m_{ch}}^{(n_s)}/2, f_{m_{ch}}^{(n_s)} + B_{m_{ch}}^{(n_s)}/2]$$

As a result, we address all islands along the vertical axis $f_1 = f_{CUT}$. A very similar reasoning can be applied to deal with the other axis, that is $f_2 = f_{CUT}$. In that case, $n_{ch} = k_{ch}$ and we address the rectangular islands defined by:

$$f_2 \in [f_{CUT} - B_{CUT}/2, f_{CUT} + B_{CUT}/2]$$
$$f_1 \in [f_{k_{ch}}^{(n_s)} - B_{k_{ch}}^{(n_s)}/2, f_{k_{ch}}^{(n_s)} + B_{k_{ch}}^{(n_s)}/2]$$

In the end, the overall integration formula reads:

$$G_{NLI}^{(n_s)}(f_{CUT}) = \frac{16}{27}\left[\gamma^{(n_s)}\right]^2 \Gamma^{(n_s)}(f_{CUT})$$

$$\left\{ G_{CUT}^{(n_s)} \sum_{\substack{n_{ch}=1 \\ n_{ch} \neq n_{CUT}^{(n_s)}}}^{N_{ch}^{(n)}} \left[G_{WDM,n_{ch}}^{(n_s)}\right]^2 \right.$$

$$\left[ \int_{f_{CUT}-B_{CUT}/2}^{f_{CUT}+B_{CUT}/2} \int_{f_{n_{ch}}^{(n_s)}-B_{n_{ch}}^{(n_s)}/2}^{f_{n_{ch}}^{(n_s)}+B_{n_{ch}}^{(n_s)}/2} |\rho(f_1,f_2,f_{CUT})|^2 df_2 df_1 + \right.$$

$$\left. \int_{f_{n_{CUT}}^{(n)}-B_{n_{CUT}}^{(n)}/2}^{f_{n_{CUT}}^{(n)}+B_{n_{CUT}}^{(n)}/2} \int_{f_{n_{ch}}^{(n_s)}-B_{n_{ch}}^{(n_s)}/2}^{f_{n_{ch}}^{(n_s)}+B_{n_{ch}}^{(n_s)}/2} |\rho(f_1,f_2,f_{CUT})|^2 df_1 df_2 \right]$$

$$\left. + \left[G_{CUT}^{(n_s)}\right]^3 \int_{f_{CUT}-B_{CUT}/2}^{f_{CUT}+B_{CUT}/2} \int_{f_{CUT}-B_{CUT}/2}^{f_{CUT}+B_{CUT}/2} |\rho(f_1,f_2,f_{CUT})|^2 df_1 df_2 \right\}$$

*Eq. 23*

The two integrals within the square brackets are very similar. One originates from the locking $n_{ch} = k_{ch}$ together with imposing $m_{ch} = n_{CUT}^{(n_s)}$; the other from the locking $n_{ch} = m_{ch}$ together with imposing $k_{ch} = n_{CUT}^{(n_s)}$. Note though that the order of integration is reversed in $f_1$ and $f_2$, in the two integrals, which are therefore apparently different. However, they would turn out to be identical if this symmetry was verified:

$$|\rho(f_1,f_2,f_{CUT})|^2 = |\rho(f_2,f_1,f_{CUT})|^2$$

Looking at Eq. 20, together with Eq. 18 and Eq. 19, this turns out to be the case. So, the two integrals are identical, can be grouped together and in the end:

$$G_{\text{NLI}}^{(n_s)}(f_{\text{CUT}}) = \frac{16}{27}\left[\gamma^{(n_s)}\right]^2 \Gamma^{(n_s)}(f_{\text{CUT}})$$

$$\left\{ 2G_{\text{CUT}}^{(n_s)} \sum_{\substack{n_{\text{ch}}=1 \\ n_{\text{ch}} \neq n_{\text{CUT}}^{(n_s)}}}^{N_{\text{ch}}^{(n)}} \left[G_{\text{WDM},n_{\text{ch}}}^{(n_s)}\right]^2 \int_{f_{\text{CUT}}-B_{\text{CUT}}/2}^{f_{\text{CUT}}+B_{\text{CUT}}/2} \int_{f_{n_{\text{ch}}}^{(n_s)}-B_{n_{\text{ch}}}^{(n_s)}/2}^{f_{n_{\text{ch}}}^{(n_s)}+B_{n_{\text{ch}}}^{(n_s)}/2} \left|\rho(f_1,f_2,f_{\text{CUT}})\right|^2 df_2 df_1 \right.$$

$$\left. + \left[G_{\text{CUT}}^{(n_s)}\right]^3 \int_{f_{\text{CUT}}-B_{\text{CUT}}/2}^{f_{\text{CUT}}+B_{\text{CUT}}/2} \int_{f_{\text{CUT}}-B_{\text{CUT}}/2}^{f_{\text{CUT}}+B_{\text{CUT}}/2} \left|\rho(f_1,f_2,f_{\text{CUT}})\right|^2 df_1 df_2 \right\}$$

*Eq. 24*

We now focus on $\left|\rho(f_1,f_2,f_{\text{CUT}})\right|^2$. From Eq. 20:

$$\left|\rho(f_1,f_2,f_{\text{CUT}})\right|^2 = \left| e^{-\int_0^{L_s^{(n_s)}} \alpha^{(n_s)}(f_{\text{CUT}},z)dz} \int_0^{L_s^{(n_s)}} e^{-j\Delta\beta^{(n_s)}(f_1,f_2,f_{\text{CUT}})z} e^{-\int_0^z \Delta\alpha^{(n_s)}(f_1,f_2,f_{\text{CUT}},z')dz'} dz \right|^2$$

This equation contains

$$\Delta\beta^{(n_s)}(f_1,f_2,f_{\text{CUT}}) = -4(f_{\text{CUT}}-f_1)(f_{\text{CUT}}-f_2)\pi^2\left(\beta_2^{(n_s)} + \pi\beta_3^{(n_s)}\left[f_1+f_2-2f_c^{(n_s)}\right]\right)$$

where it is apparent the presence of the product:

$$(f_{\text{CUT}}-f_1)(f_{\text{CUT}}-f_2)$$

As a strategy for simplifying this factor, we are going to change integration variables into:

$$f_1 = f_1' + f_{n_{\text{CUT}}}^{(n)}$$
$$f_2 = f_2' + f_{n_{\text{CUT}}}^{(n)}$$

Doing this we first get:

$$G_{\text{NLI}}^{(n_s)}(f_{\text{CUT}}) = \frac{16}{27}\left[\gamma^{(n_s)}\right]^2 \Gamma^{(n_s)}(f_{\text{CUT}})$$

$$\left\{ 2G_{\text{CUT}}^{(n_s)} \sum_{\substack{n_{\text{ch}}=1 \\ n_{\text{ch}} \neq n_{\text{CUT}}^{(n_s)}}}^{N_{\text{ch}}^{(n)}} \left[G_{\text{WDM},n_{\text{ch}}}^{(n_s)}\right]^2 \int_{-B_{\text{CUT}}/2}^{B_{\text{CUT}}/2} \int_{f_{n_{\text{ch}}}^{(n_s)}-f_{\text{CUT}}-B_{n_{\text{ch}}}^{(n_s)}/2}^{f_{n_{\text{ch}}}^{(n_s)}-f_{\text{CUT}}+B_{n_{\text{ch}}}^{(n_s)}/2} \left|\rho(f_1'+f_{\text{CUT}},f_2'+f_{\text{CUT}},f_{\text{CUT}})\right|^2 df_2' df_1' \right.$$

$$\left. + \left[G_{\text{CUT}}^{(n_s)}\right]^3 \int_{f_{\text{CUT}}-B_{\text{CUT}}/2}^{f_{\text{CUT}}+B_{\text{CUT}}/2} \int_{f_{\text{CUT}}-B_{\text{CUT}}/2}^{f_{\text{CUT}}+B_{\text{CUT}}/2} \left|\rho(f_1'+f_{\text{CUT}},f_2'+f_{\text{CUT}},f_{\text{CUT}})\right|^2 df_1' df_2' \right\}$$

*Eq. 25*

and:

$$\left|\rho\left(f_1' + f_{\text{CUT}}, f_2' + f_{\text{CUT}}, f_{\text{CUT}}\right)\right|^2$$

$$= \left|e^{-\int_0^{L_s^{(n_s)}} \alpha^{(n_s)}(f_{\text{CUT}},z)dz} L_s^{(n_s)} \int_0^{L_s^{(n_s)}} e^{-j\Delta\beta^{(n_s)}\left(f_1'+f_{\text{CUT}},f_2'+f_{\text{CUT}},f_{\text{CUT}}\right)z} e^{-\int_0^z \Delta\alpha^{(n_s)}\left(f_1'+f_{\text{CUT}},f_2'+f_{\text{CUT}},f_{\text{CUT}},z'\right)dz'} dz\right|^2$$

and, as a consequence:

$$\Delta\beta^{(n_s)}\left(f_1' + f_{\text{CUT}}, f_2' + f_{\text{CUT}}, f_{\text{CUT}}\right) = -4\pi^2 f_1' f_2' \left(\beta_2^{(n_s)} + \pi\beta_3^{(n_s)}\left[f_1' + f_2' + 2f_{\text{CUT}} - 2f_c^{(n_s)}\right]\right)$$

Performing the same substitution on the loss-related term:

$$\Delta\alpha^{(n_s)}\left(f_1' + f_{\text{CUT}}, f_2' + f_{\text{CUT}}, f_{\text{CUT}}, z\right) =$$

$$= \alpha^{(n_s)}\left(f_1' + f_{\text{CUT}}, z\right) + \alpha^{(n_s)}\left(f_2' + f_{\text{CUT}}, z\right) - \alpha^{(n_s)}\left(f_{\text{CUT}}, z\right) + \alpha^{(n_s)}\left(f_1' + f_{\text{CUT}} + f_2' + f_{\text{CUT}} - f_{\text{CUT}}, z\right)$$

$$= \alpha^{(n_s)}\left(f_1' + f_{\text{CUT}}, z\right) + \alpha^{(n_s)}\left(f_2' + f_{\text{CUT}}, z\right) - \alpha^{(n_s)}\left(f_{\text{CUT}}, z\right) + \alpha^{(n_s)}\left(f_1' + f_2' + f_{\text{CUT}}, z\right)$$

$$= \alpha_0^{(n_s)}\left(f_1' + f_{\text{CUT}}\right) + \alpha_1^{(n_s)}\left(f_1' + f_{\text{CUT}}\right)\exp\left(-\sigma^{(n_s)}\left(f_1' + f_{\text{CUT}}\right)\cdot z\right) +$$

$$\alpha_0^{(n_s)}\left(f_2' + f_{\text{CUT}}\right) + \alpha_1^{(n_s)}\left(f_2' + f_{\text{CUT}}\right)\exp\left(-\sigma^{(n_s)}\left(f_2' + f_{\text{CUT}}\right)\cdot z\right) -$$

$$\alpha_0^{(n_s)}\left(f_{\text{CUT}}\right) + \alpha_1^{(n_s)}\left(f_{\text{CUT}}\right)\exp\left(-\sigma^{(n_s)}\left(f_{\text{CUT}}\right)\cdot z\right) +$$

$$\alpha_0^{(n_s)}\left(f_1' + f_2' + f_{\text{CUT}}\right) + \alpha_1^{(n_s)}\left(f_1' + f_2' + f_{\text{CUT}}\right)\exp\left(-\sigma^{(n_s)}\left(f_1' + f_2' + f_{\text{CUT}}\right)\cdot z\right)$$

We are now going to take a few *fundamental approximations.*

Specifically, in some of the terms above, we are going to substitute $f_1'$ and $f_1'$ with their average values, that is with 0 and $\left(f_{n_{\text{ch}}}^{(n_s)} - f_{\text{CUT}}\right)$, respectively. We get:

$$\Delta\beta^{(n_s)}\left(f_1' + f_{\text{CUT}}, f_2' + f_{\text{CUT}}, f_{\text{CUT}}\right)$$

$$= -4\pi^2 f_1' f_2' \left(\beta_2^{(n_s)} + \pi\beta_3^{(n_s)}\left[f_1' + f_2' + 2f_{\text{CUT}} - 2f_c^{(n_s)}\right]\right)$$

$$\approx -4\pi^2 f_1' f_2' \left(\beta_2^{(n_s)} + \pi\beta_3^{(n_s)}\left[f_{n_{\text{ch}}}^{(n_s)} - f_{\text{CUT}} + 2f_{\text{CUT}} - 2f_c^{(n_s)}\right]\right)$$

$$= -4\pi^2 f_1' f_2' \left(\beta_2^{(n_s)} + \pi\beta_3^{(n_s)}\left[f_{n_{\text{ch}}}^{(n_s)} + f_{\text{CUT}} - 2f_c^{(n_s)}\right]\right)$$

which can then be re-written as follows:

$$\Delta\beta^{(n_s)}\left(f_1' + f_{\text{CUT}}, f_2' + f_{\text{CUT}}, f_{\text{CUT}}\right) \approx -4\pi^2 f_1' f_2' \cdot \beta_{2\text{eff},n_{\text{ch}}}^{(n_s)}$$

where:

$$\beta_{2\text{eff},n_{\text{ch}}}^{(n_s)} = \beta_2^{(n_s)} + \pi\beta_3^{(n_s)}\left[f_{n_{\text{ch}}}^{(n_s)} + f_{\text{CUT}} - 2f_c^{(n_s)}\right]$$

and, for the CUT:

$$\beta^{(n_s)}_{2\text{eff},n^{(n_s)}_{\text{CUT}}} = \beta^{(n_s)}_2 + \pi\beta^{(n_s)}_3 \left[ 2f_{\text{CUT}} - 2f^{(n_s)}_c \right]$$

To better understand the meaning of this approximation on $\Delta\beta^{(n_s)}$, we point out that the result would be exact if $\beta_2$ was *piecewise constant* over the bandwidth of each channel. So, we are indeed accounting for the frequency change of dispersion, but we assume that it is locally constant over each channel.

We then apply the same strategy to the loss term and again we obtain a dramatic simplification:

$$\Delta\alpha^{(n_s)}\left(f'_1 + f_{\text{CUT}}, f'_2 + f_{\text{CUT}}, f_{\text{CUT}}, z\right) \approx$$

$$\alpha^{(n_s)}_0\left(f_{\text{CUT}}\right) + \alpha^{(n_s)}_1\left(f_{\text{CUT}}\right)\exp\left(-\sigma^{(n_s)}\left(f_{\text{CUT}}\right)\cdot z\right) +$$

$$\alpha^{(n_s)}_0\left(f^{(n_s)}_{n_{\text{ch}}} - f_{\text{CUT}} + f_{\text{CUT}}\right) + \alpha^{(n_s)}_1\left(f^{(n_s)}_{n_{\text{ch}}} - f_{\text{CUT}} + f_{\text{CUT}}\right)\exp\left(-\sigma^{(n_s)}\left(f^{(n_s)}_{n_{\text{ch}}} - f_{\text{CUT}} + f_{\text{CUT}}\right)\cdot z\right) -$$

$$\alpha^{(n_s)}_0\left(f_{\text{CUT}}\right) + \alpha^{(n_s)}_1\left(f_{\text{CUT}}\right)\exp\left(-\sigma^{(n_s)}\left(f_{\text{CUT}}\right)\cdot z\right) +$$

$$\alpha^{(n_s)}_0\left(f^{(n_s)}_{n_{\text{ch}}} - f_{\text{CUT}} + f_{\text{CUT}}\right) + \alpha^{(n_s)}_1\left(f^{(n_s)}_{n_{\text{ch}}} - f_{\text{CUT}} + f_{\text{CUT}}\right)\exp\left(-\sigma^{(n_s)}\left(f^{(n_s)}_{n_{\text{ch}}} - f_{\text{CUT}} + f_{\text{CUT}}\right)\cdot z\right)$$

$$= 2\alpha^{(n_s)}_0\left(f^{(n_s)}_{n_{\text{ch}}}\right) + 2\alpha^{(n_s)}_1\left(f^{(n_s)}_{n_{\text{ch}}}\right)\exp\left(-\sigma^{(n_s)}\left(f^{(n_s)}_{n_{\text{ch}}}\right)\cdot z\right)$$

Once more, the result would be exact if loss was frequency-flat over the bandwidth of each of the channels involved. Quite remarkably, under this approximation, it appears that what matters is loss at a single channel (the one interfering with the CUT to produce XCI) and nowhere else.

We can then insert these results into $\rho$:

$$\left|\rho\left(f'_1 + f_{\text{CUT}}, f'_2 + f_{\text{CUT}}, f_{\text{CUT}}\right)\right|^2 =$$

$$\left| e^{-\int_0^{L^{(n_s)}_s} \alpha^{(n_s)}(f_{\text{CUT}},z)dz} \int_0^{L^{(n_s)}_s} e^{-j\Delta\beta^{(n_s)}\left(f'_1 + f_{\text{CUT}}, f'_2 + f_{\text{CUT}}, f_{\text{CUT}}\right)z} e^{-\int_0^z \Delta\alpha^{(n_s)}\left(f'_1 + f_{\text{CUT}}, f'_2 + f_{\text{CUT}}, f_{\text{CUT}}, z'\right)dz'} dz \right|^2 =$$

$$\left| e^{-\int_0^{L^{(n_s)}_s} \left[\alpha^{(n_s)}_0(f_{\text{CUT}}) + \alpha^{(n_s)}_1(f_{\text{CUT}})\exp\left(-\sigma^{(n_s)}(f_{\text{CUT}})\cdot z\right)\right]dz} \int_0^{L^{(n_s)}_s} e^{j4\pi^2 f'_1 f'_2 \cdot \beta^{(n_s)}_{2\text{eff},n_{\text{ch}}} z} e^{-2\int_0^z \left[\alpha^{(n_s)}_0\left(f^{(n_s)}_{n_{\text{ch}}}\right) + \alpha^{(n_s)}_1\left(f^{(n_s)}_{n_{\text{ch}}}\right)\exp\left(-\sigma^{(n_s)}\left(f^{(n_s)}_{n_{\text{ch}}}\right)\cdot z'\right)\right]dz'} dz \right|^2$$

*Eq. 26*

### 4. Integration

We now confront with the task of carrying out one spatial integration and two frequency integrations, cascaded. We first look at the spatial integrals within the factor $\rho$. One of the integrals has the form:

$$\int_0^L e^{-jf'_1 f'_2 \cdot Bz} e^{-2\int_0^z \left[\alpha_0 + \alpha_1 \exp(-\sigma z')\right]dz'} dz$$

This integral cannot be executed exactly. To get convergence, it is necessary to assume that the integration limit be infinity, instead of the span length. We accept such approximation and as a result we get:

$$\int_0^\infty e^{-jf_1'f_2'\cdot Bz} e^{-2\int_0^z [\alpha_0+\alpha_1 \exp(-\sigma z')]dz'} dz$$

$$= \frac{e^{-\frac{2\alpha_1}{\sigma}} \left(-\frac{2\alpha_1}{\sigma}\right)^{-\frac{2\alpha_0+jf_1'f_2'\cdot B}{\sigma}} \left(\Gamma\left(\frac{2\alpha_0+jf_1'f_2'\cdot B}{\sigma}\right) - \Gamma\left(\frac{2\alpha_0+jf_1'f_2'\cdot B}{\sigma}, -\frac{2\alpha_1}{\sigma}\right)\right)}{\sigma}$$

We point out that in virtual totality of practical cases this approximation is immaterial. It amounts to summing NLI produced over in ideally infinitely long fiber, rather than one $L_s^{(n_s)}$ km long. Since fibers are lossy, by far most of NLI is actually produced over the first 20-30 km of fiber, and virtually all within the first 50 km, unless some form of distributed amplification is used. Even so, this approximation is likely not to be a problem, but we leave the discussion of this case for a subsequent paper.

There is also an outer integral, which can be executed exactly:

$$e^{-\int_0^L [\alpha_0+\alpha_1\exp(-\sigma z)]dz} = e^{\frac{\alpha_1(e^{-\sigma L}-1)}{\sigma}-\alpha_0 L} = e^{-\alpha_0 L}e^{\alpha_1(e^{-\sigma L}-1)/\sigma}$$

Substituting the results so far, we get:

$$\left|\rho\left(f_1'+f_{\text{CUT}}, f_2'+f_{\text{CUT}}, f_{\text{CUT}}\right)\right|^2 =$$

$$\left| e^{-\int_0^{L_s^{(n_s)}}\left[\alpha_0^{(n_s)}(f_{\text{CUT}})+\alpha_1^{(n_s)}(f_{\text{CUT}})\exp\left(-\sigma^{(n_s)}(f_{\text{CUT}})\cdot z\right)\right]dz} \int_0^{L_s^{(n_s)}} e^{j4\pi^2 f_1' f_2' \cdot \beta_{2\text{eff},n_{\text{ch}}}^{(n_s)} z} e^{-2\int_0^z\left[\alpha_0^{(n_s)}\left(f_{n_{\text{ch}}}^{(n_s)}\right)+\alpha_1^{(n_s)}\left(f_{n_{\text{ch}}}^{(n_s)}\right)\exp\left(-\sigma^{(n_s)}\left(f_{n_{\text{ch}}}^{(n_s)}\right)\cdot z'\right)\right]dz'} dz \right|^2$$

$$= e^{-2\alpha_0^{(n_s)}(f_{\text{CUT}})L_s^{(n_s)}} e^{2\alpha_1^{(n_s)}(f_{\text{CUT}})\left(e^{-\sigma^{(n_s)}(f_{\text{CUT}})L_s^{(n_s)}}-1\right)/\sigma^{(n_s)}(f_{\text{CUT}})}$$

$$\frac{1}{\left[\sigma^{(n_s)}\left(f_{n_{\text{ch}}}^{(n_s)}\right)\right]^2} \left| e^{-\frac{2\alpha_1^{(n_s)}\left(f_{n_{\text{ch}}}^{(n_s)}\right)}{\sigma^{(n_s)}\left(f_{n_{\text{ch}}}^{(n_s)}\right)}} \left(-\frac{2\alpha_1^{(n_s)}\left(f_{n_{\text{ch}}}^{(n_s)}\right)}{\sigma^{(n_s)}\left(f_{n_{\text{ch}}}^{(n_s)}\right)}\right)^{-\frac{2\alpha_0^{(n_s)}\left(f_{n_{\text{ch}}}^{(n_s)}\right)-jf_1'f_2'\cdot 4\pi^2 \beta_{2\text{eff},n_{\text{ch}}}^{(n_s)}}{\sigma^{(n_s)}\left(f_{n_{\text{ch}}}^{(n_s)}\right)}} \right|^2$$

$$\left|\left(\Gamma\left(\frac{2\alpha_0^{(n_s)}\left(f_{n_{\text{ch}}}^{(n_s)}\right)-jf_1'f_2'\cdot 4\pi^2 \beta_{2\text{eff},n_{\text{ch}}}^{(n_s)}}{\sigma^{(n_s)}\left(f_{n_{\text{ch}}}^{(n_s)}\right)}\right) - \Gamma\left(\frac{2\alpha_0^{(n_s)}\left(f_{n_{\text{ch}}}^{(n_s)}\right)-jf_1'f_2'\cdot 4\pi^2 \beta_{2\text{eff},n_{\text{ch}}}^{(n_s)}}{\sigma^{(n_s)}\left(f_{n_{\text{ch}}}^{(n_s)}\right)}, -\frac{2\alpha_1^{(n_s)}\left(f_{n_{\text{ch}}}^{(n_s)}\right)}{\sigma^{(n_s)}\left(f_{n_{\text{ch}}}^{(n_s)}\right)}\right)\right)\right|^2$$

*Eq. 27*

For the sake of readability, we are going to replace in some of the next few formulas:

$$\alpha_1^{(n_s)}\left(f_{n_{ch}}^{(n_s)}\right) \to \alpha_1$$

$$\alpha_0^{(n_s)}\left(f_{n_{ch}}^{(n_s)}\right) \to \alpha_0$$

$$\sigma^{(n_s)}\left(f_{n_{ch}}^{(n_s)}\right) \to \sigma$$

$$4\pi^2 \beta_{2\text{eff},n_{ch}}^{(n_s)} \to B$$

Therefore:

$$\left|\rho\left(f_1' + f_{\text{CUT}}, f_2' + f_{\text{CUT}}, f_{\text{CUT}}\right)\right|^2 =$$

$$= e^{-2\alpha_0 L_s^{(n_s)}} e^{2\alpha_1^{(n_s)}(f_{\text{CUT}})\left(e^{-\sigma^{(n_s)}(f_{\text{CUT}})L}-1\right)/\sigma^{(n_s)}(f_{\text{CUT}})}$$

$$\frac{1}{\sigma^2}\left|e^{-\frac{2\alpha_1}{\sigma}}\left(-\frac{2\alpha_1}{\sigma}\right)^{-\frac{2\alpha_0-jf_1'f_2'\cdot B}{\sigma}}\left(\Gamma\left(\frac{2\alpha_0-jf_1'f_2'\cdot B}{\sigma}\right)-\Gamma\left(\frac{2\alpha_0-jf_1'f_2'\cdot B}{\sigma},-\frac{2\alpha_1}{\sigma}\right)\right)\right|^2$$

We can then go back to Eq. 25. We can re-write it as:

$$G_{\text{NLI}}^{(n_s)}\left(f_{\text{CUT}}\right) = \frac{16}{27}\left[\gamma^{(n_s)}\right]^2 \Gamma^{(n_s)}\left(f_{\text{CUT}}\right) e^{-2\alpha_0^{(n_s)}(f_{\text{CUT}})\cdot L_s^{(n_s)}} e^{2\alpha_1^{(n_s)}(f_{\text{CUT}})\left(e^{-\sigma^{(n_s)}(f_{\text{CUT}})L_s^{(n_s)}}-1\right)/\sigma^{(n_s)}(f_{\text{CUT}})}$$

$$\left\{ 2\frac{G_{\text{CUT}}^{(n_s)}}{\left[\sigma^{(n_s)}\left(f_{n_{ch}}^{(n_s)}\right)\right]^2} \sum_{\substack{n_{ch}=1 \\ n_{ch} \neq n_{\text{CUT}}^{(n_s)}}}^{N_{ch}^{(n)}} \left[G_{\text{WDM},n_{ch}}^{(n_s)}\right]^2 \int_{-B_{\text{CUT}}/2}^{B_{\text{CUT}}/2} \int_{f_{n_{ch}}^{(n_s)}-f_{\text{CUT}}-B_{n_{ch}}^{(n_s)}/2}^{f_{n_{ch}}^{(n_s)}-f_{\text{CUT}}+B_{n_{ch}}^{(n_s)}/2} \right.$$

$$\left|e^{-\frac{2\alpha_1^{(n_s)}\left(f_{n_{ch}}^{(n_s)}\right)}{\sigma^{(n_s)}\left(f_{n_{ch}}^{(n_s)}\right)}}\left(-\frac{2\alpha_1^{(n_s)}\left(f_{n_{ch}}^{(n_s)}\right)}{\sigma^{(n_s)}\left(f_{n_{ch}}^{(n_s)}\right)}\right)^{-\frac{2\alpha_0^{(n_s)}\left(f_{n_{ch}}^{(n_s)}\right)-jf_1'f_2'\cdot 4\pi^2 \beta_{2\text{eff},n_{ch}}^{(n_s)}}{\sigma^{(n_s)}\left(f_{n_{ch}}^{(n_s)}\right)}}\right|^2$$

$$\left|\left(\Gamma\left(\frac{2\alpha_0^{(n_s)}\left(f_{n_{ch}}^{(n_s)}\right)-jf_1'f_2'\cdot 4\pi^2 \beta_{2\text{eff},n_{ch}}^{(n_s)}}{\sigma^{(n_s)}\left(f_{n_{ch}}^{(n_s)}\right)}\right)-\Gamma\left(\frac{2\alpha_0^{(n_s)}\left(f_{n_{ch}}^{(n_s)}\right)-jf_1'f_2'\cdot 4\pi^2 \beta_{2\text{eff},n_{ch}}^{(n_s)}}{\sigma^{(n_s)}\left(f_{n_{ch}}^{(n_s)}\right)},-\frac{2\alpha_1^{(n_s)}\left(f_{n_{ch}}^{(n_s)}\right)}{\sigma^{(n_s)}\left(f_{n_{ch}}^{(n_s)}\right)}\right)\right)\right|^2 df_2' df_1'$$

$$+\frac{\left[G_{\text{CUT}}^{(n_s)}\right]^3}{\left[\sigma^{(n_s)}\left(f_{\text{CUT}}\right)\right]^2} \int_{f_{\text{CUT}}-B_{\text{CUT}}/2}^{f_{\text{CUT}}+B_{\text{CUT}}/2} \int_{f_{\text{CUT}}-B_{\text{CUT}}/2}^{f_{\text{CUT}}+B_{\text{CUT}}/2} \left|e^{-\frac{2\alpha_1^{(n_s)}(f_{\text{CUT}})}{\sigma^{(n_s)}(f_{\text{CUT}})}}\left(-\frac{2\alpha_1^{(n_s)}(f_{\text{CUT}})}{\sigma^{(n_s)}(f_{\text{CUT}})}\right)^{-\frac{2\alpha_0^{(n_s)}(f_{\text{CUT}})-jf_1'f_2'\cdot 4\pi^2 \beta_{2\text{eff},n_{\text{CUT}}^{(n_s)}}^{(n_s)}}{\sigma^{(n_s)}(f_{\text{CUT}})}}\right|^2$$

$$\left|\left(\Gamma\left(\frac{2\alpha_0^{(n_s)}\left(f_{\text{CUT}}\right)-jf_1'f_2'\cdot 4\pi^2 \beta_{2\text{eff},n_{\text{CUT}}^{(n_s)}}^{(n_s)}}{\sigma^{(n_s)}\left(f_{\text{CUT}}\right)}\right)-\Gamma\left(\frac{2\alpha_0^{(n_s)}\left(f_{\text{CUT}}\right)-jf_1'f_2'\cdot 4\pi^2 \beta_{2\text{eff},n_{\text{CUT}}^{(n_s)}}^{(n_s)}}{\sigma^{(n_s)}\left(f_{\text{CUT}}\right)},-\frac{2\alpha_1^{(n_s)}\left(f_{\text{CUT}}\right)}{\sigma^{(n_s)}\left(f_{\text{CUT}}\right)}\right)\right)\right|^2 df_1' df_2' \right\}$$

*Eq. 28*

The formula can then be compacted as follows:

$$G_{\text{NLI}}^{(n_s)}(f_{\text{CUT}}) = \frac{16}{27}\left[\gamma^{(n_s)}\right]^2 \Gamma^{(n_s)}(f_{\text{CUT}}) e^{-2\alpha_0^{(n_s)}(f_{\text{CUT}})\cdot L_s^{(n_s)}} e^{2\alpha_1^{(n_s)}(f_{\text{CUT}})\left(e^{-\sigma^{(n_s)}(f_{\text{CUT}})L_s^{(n_s)}}-1\right)/\sigma^{(n_s)}(f_{\text{CUT}})}$$

$$G_{\text{CUT}}^{(n_s)}\left[\sum_{\substack{n_{\text{ch}}=1 \\ n_{\text{ch}} \neq n_{\text{CUT}}}}^{N_{\text{ch}}^{(n)}} \left[G_{\text{WDM},n_{\text{ch}}}^{(n_s)}\right]^2 2I_{n_{\text{ch}}}^{(n_s)} + \left[G_{\text{CUT}}^{(n_s)}\right]^2 I_{n_{\text{CUT}}}^{(n_s)}\right]$$

*Eq. 29*

where:

$$I_{n_{\text{ch}}}^{(n_s)} = \frac{1}{\left[\sigma^{(n_s)}\left(f_{n_{\text{ch}}}^{(n_s)}\right)\right]^2} \int_{-B_{\text{CUT}}/2}^{B_{\text{CUT}}/2} \int_{f_{n_{\text{ch}}}^{(n_s)}-f_{\text{CUT}}-B_{n_{\text{ch}}}^{(n_s)}/2}^{f_{n_{\text{ch}}}^{(n_s)}-f_{\text{CUT}}+B_{n_{\text{ch}}}^{(n_s)}/2} \left| e^{-\frac{2\alpha_1^{(n_s)}\left(f_{n_{\text{ch}}}^{(n_s)}\right)}{\sigma^{(n_s)}\left(f_{n_{\text{ch}}}^{(n_s)}\right)}} \left(-\frac{2\alpha_1^{(n_s)}\left(f_{n_{\text{ch}}}^{(n_s)}\right)}{\sigma^{(n_s)}\left(f_{n_{\text{ch}}}^{(n_s)}\right)}\right)^{-\frac{2\alpha_0^{(n_s)}\left(f_{n_{\text{ch}}}^{(n_s)}\right)-jf_1'f_2'\cdot 4\pi^2\beta_{2\text{eff},n_{\text{ch}}}^{(n_s)}}{\sigma^{(n_s)}\left(f_{n_{\text{ch}}}^{(n_s)}\right)}} \right|^2$$

$$\left| \left(\Gamma\left(\frac{2\alpha_0^{(n_s)}\left(f_{n_{\text{ch}}}^{(n_s)}\right)-jf_1'f_2'\cdot 4\pi^2\beta_{2\text{eff},n_{\text{ch}}}^{(n_s)}}{\sigma^{(n_s)}\left(f_{n_{\text{ch}}}^{(n_s)}\right)}\right) - \Gamma\left(\frac{2\alpha_0^{(n_s)}\left(f_{n_{\text{ch}}}^{(n_s)}\right)-jf_1'f_2'\cdot 4\pi^2\beta_{2\text{eff},n_{\text{ch}}}^{(n_s)}}{\sigma^{(n_s)}\left(f_{n_{\text{ch}}}^{(n_s)}\right)}, -\frac{2\alpha_1^{(n_s)}\left(f_{n_{\text{ch}}}^{(n_s)}\right)}{\sigma^{(n_s)}\left(f_{n_{\text{ch}}}^{(n_s)}\right)}\right)\right) \right|^2 df_2' df_1'$$

*Eq. 30*

$$I_{\text{CUT}}^{(n_s)} = \frac{1}{\left[\sigma^{(n_s)}(f_{\text{CUT}})\right]^2} \int_{f_{\text{CUT}}-B_{\text{CUT}}/2}^{f_{\text{CUT}}+B_{\text{CUT}}/2} \int_{f_{\text{CUT}}-B_{\text{CUT}}/2}^{f_{\text{CUT}}+B_{\text{CUT}}/2} \left| e^{-\frac{2\alpha_1^{(n_s)}(f_{\text{CUT}})}{\sigma^{(n_s)}(f_{\text{CUT}})}} \left(-\frac{2\alpha_1^{(n_s)}(f_{\text{CUT}})}{\sigma^{(n_s)}(f_{\text{CUT}})}\right)^{-\frac{2\alpha_0^{(n_s)}(f_{\text{CUT}})-jf_1'f_2'\cdot 4\pi^2\beta_{2\text{eff},n_{\text{CUT}}}^{(n_s)}}{\sigma^{(n_s)}(f_{\text{CUT}})}} \right|^2$$

$$\left| \left(\Gamma\left(\frac{2\alpha_0^{(n_s)}(f_{\text{CUT}})-jf_1'f_2'\cdot 4\pi^2\beta_{2\text{eff},n_{\text{CUT}}}^{(n_s)}}{\sigma^{(n_s)}(f_{\text{CUT}})}\right) - \Gamma\left(\frac{2\alpha_0^{(n_s)}(f_{\text{CUT}})-jf_1'f_2'\cdot 4\pi^2\beta_{2\text{eff},n_{\text{CUT}}}^{(n_s)}}{\sigma^{(n_s)}(f_{\text{CUT}})}, -\frac{2\alpha_1^{(n_s)}(f_{\text{CUT}})}{\sigma^{(n_s)}(f_{\text{CUT}})}\right)\right) \right|^2 df_1' df_2' \}$$

*Eq. 31*

The first integral outlined above, $I_{n_{\text{ch}}}^{(n_s)}$ in Eq. 30, has the notationally simplified form:

$$I_{n_{\text{ch}}}^{(n_s)} = \int_{-B_{\text{CUT}}/2}^{B_{\text{CUT}}/2} \int_{f_{n_{\text{ch}}}^{(n_s)}-f_{\text{CUT}}-B_{n_{\text{ch}}}^{(n_s)}/2}^{f_{n_{\text{ch}}}^{(n_s)}-f_{\text{CUT}}+B_{n_{\text{ch}}}^{(n_s)}/2}$$

$$\left| e^{-\frac{2\alpha_1}{\sigma}} \left(-\frac{2\alpha_1}{\sigma}\right)^{-\frac{2\alpha_0-jf_1'f_2'\cdot B}{\sigma}} \left(\Gamma\left(\frac{2\alpha_0-jf_1'f_2'\cdot B}{\sigma}\right) - \Gamma\left(\frac{2\alpha_0-jf_1'f_2'\cdot B}{\sigma}, -\frac{2\alpha_1}{\sigma}\right)\right) \right|^2 df_1' df_2'$$

Unfortunately, integration in frequency does not appear to be possible in closed form. We have then to take some further approximations. The key assumption that we make is:

$\alpha_1 \ll \alpha_0$

In other words, the presence of an SRS-induced $\alpha_1$ is assumed to be a small perturbation vs. the bulk of loss, which is due to fiber loss $\alpha_0$. This appears to be reasonable. If so, we can think of replacing the integrand with a series expansion of it. In particular, we take the expression inside the absolute value squared and expand it first order vs. $\alpha_1$, about zero. We get:

$$e^{-\frac{2\alpha_1}{\sigma}}\left(-\frac{2\alpha_1}{\sigma}\right)^{-\frac{2\alpha_0 - jf_1'f_2' \cdot B}{\sigma}}\left(\Gamma\left(\frac{2\alpha_0 - jf_1'f_2' \cdot B}{\sigma}\right) - \Gamma\left(\frac{2\alpha_0 - jf_1'f_2' \cdot B}{\sigma}, -\frac{2\alpha_1}{\sigma}\right)\right) =$$

$$\frac{1}{2\alpha_0 - jf_1'f_2' \cdot B} - \frac{2\alpha_1}{(2\alpha_0 - jf_1'f_2' \cdot B)(2\alpha_0 - jf_1'f_2' \cdot B + \sigma)} + O(\alpha_1^2)$$

Remarkably, the resulting expression is free of special functions. This expression needs then to be absolute-value squared. The calculation yields:

$$\left|\frac{1}{2\alpha_0 - jf_1'f_2' \cdot B} - \frac{2\alpha_1}{(2\alpha_0 - jf_1'f_2' \cdot B)(2\alpha_0 - jf_1'f_2' \cdot B + \sigma)}\right|^2 =$$

$$\left[\frac{1}{2\alpha_0 - jf_1'f_2' \cdot B} - \frac{2\alpha_1}{(2\alpha_0 - jf_1'f_2' \cdot B)(2\alpha_0 - jf_1'f_2' \cdot B + \sigma)}\right] \cdot$$

$$\left[\frac{1}{2\alpha_0 + jf_1'f_2' \cdot B} - \frac{2\alpha_1}{(2\alpha_0 + jf_1'f_2' \cdot B)(2\alpha_0 + jf_1'f_2' \cdot B + \sigma)}\right]$$

$$= \frac{(2\alpha_0 - 2\alpha_1 - jf_1'f_2' \cdot B + \sigma)(2\alpha_0 - 2\alpha_1 + jf_1'f_2' \cdot B + \sigma)}{(4\alpha_0^2 + f_1'^2 f_2'^2 \cdot B^2)((2\alpha_0 + \sigma)^2 + f_1'^2 f_2'^2 \cdot B^2)}$$

$$= \frac{(2\alpha_0 - 2\alpha_1 + \sigma)^2 + f_1'^2 f_2'^2 \cdot B^2}{(4\alpha_0^2 + f_1'^2 f_2'^2 \cdot B^2)((2\alpha_0 + \sigma)^2 + f_1'^2 f_2'^2 \cdot B^2)}$$

We can then put it into $I_{n_{\text{ch}}}^{(n_s)}$, in Eq. 30:

$$I_{n_{\text{ch}}}^{(n_s)} = \int_{-B_{\text{CUT}}/2}^{B_{\text{CUT}}/2} \int_{f_{n_{\text{ch}}}^{(n_s)} - f_{\text{CUT}} - B_{n_{\text{ch}}}^{(n_s)}/2}^{f_{n_{\text{ch}}}^{(n_s)} - f_{\text{CUT}} + B_{n_{\text{ch}}}^{(n_s)}/2} \frac{(2\alpha_0 - 2\alpha_1 + \sigma)^2 + f_1'^2 f_2'^2 \cdot B^2}{(4\alpha_0^2 + f_1'^2 f_2'^2 \cdot B^2)((2\alpha_0 + \sigma)^2 + f_1'^2 f_2'^2 \cdot B^2)} df_1' df_2'$$

Quite remarkably this double integral can be carried out in *fully analytical closed form*:

$$I_{n_{ch}}^{(n_s)} = -\frac{j}{2\alpha_0 B\sigma(2\alpha_0+\sigma)(4\alpha_0+\sigma)}\{(2\alpha_0+\sigma)(-2\alpha_1+\sigma)(4\alpha_0-2\alpha_1+\sigma)$$

$$[\;\text{Li}_2\left(-j\left|\frac{B}{2\alpha_0}\right|\left[f_{n_{ch}}^{(n_s)}-f_{\text{CUT}}-\frac{B_{n_{ch}}^{(n_s)}}{2}\right]\frac{B_{\text{CUT}}}{2}\right)-\text{Li}_2\left(j\left|\frac{B}{2\alpha_0}\right|\left[f_{n_{ch}}^{(n_s)}-f_{\text{CUT}}-\frac{B_{n_{ch}}^{(n_s)}}{2}\right]\frac{B_{\text{CUT}}}{2}\right)$$

$$-\text{Li}_2\left(-j\left|\frac{B}{2\alpha_0}\right|\left[f_{n_{ch}}^{(n_s)}-f_{\text{CUT}}+\frac{B_{n_{ch}}^{(n_s)}}{2}\right]\frac{B_{\text{CUT}}}{2}\right)+\text{Li}_2\left(j\left|\frac{B}{2\alpha_0}\right|\left[f_{n_{ch}}^{(n_s)}-f_{\text{CUT}}+\frac{B_{n_{ch}}^{(n_s)}}{2}\right]\frac{B_{\text{CUT}}}{2}\right)\;]$$

$$+8\alpha_0\alpha_1(2\alpha_0-\alpha_1+\sigma)$$

$$[\;\text{Li}_2\left(-j\left|\frac{B}{2\alpha_0+\sigma}\right|\left[f_{n_{ch}}^{(n_s)}-f_{\text{CUT}}-\frac{B_{n_{ch}}^{(n_s)}}{2}\right]\frac{B_{\text{CUT}}}{2}\right)-\text{Li}_2\left(j\left|\frac{B}{2\alpha_0+\sigma}\right|\left[f_{n_{ch}}^{(n_s)}-f_{\text{CUT}}-\frac{B_{n_{ch}}^{(n_s)}}{2}\right]\frac{B_{\text{CUT}}}{2}\right)$$

$$-\text{Li}_2\left(-j\left|\frac{B}{2\alpha_0+\sigma}\right|\left[f_{n_{ch}}^{(n_s)}-f_{\text{CUT}}+\frac{B_{n_{ch}}^{(n_s)}}{2}\right]\frac{B_{\text{CUT}}}{2}\right)+\text{Li}_2\left(j\left|\frac{B}{2\alpha_0+\sigma}\right|\left[f_{n_{ch}}^{(n_s)}-f_{\text{CUT}}+\frac{B_{n_{ch}}^{(n_s)}}{2}\right]\frac{B_{\text{CUT}}}{2}\right)\;]\;\}$$

Similarly, for

$$I_{\text{CUT}}^{(n_s)} = \frac{j}{\alpha_0 B\sigma(2\alpha_0+\sigma)(4\alpha_0+\sigma)}\{(2\alpha_0+\sigma)(-2\alpha_1+\sigma)(4\alpha_0-2\alpha_1+\sigma)$$

$$[\;\text{Li}_2\left(-j\left|\frac{B}{2\alpha_0}\right|\frac{B_{\text{CUT}}^2}{4}\right)-\text{Li}_2\left(j\left|\frac{B}{2\alpha_0}\right|\frac{B_{\text{CUT}}^2}{4}\right)\;]$$

$$+8\alpha_0\alpha_1(2\alpha_0-\alpha_1+\sigma)$$

$$[\;\text{Li}_2\left(-j\left|\frac{B}{2\alpha_0+\sigma}\right|\frac{B_{\text{CUT}}^2}{4}\right)-\text{Li}_2\left(j\left|\frac{B}{2\alpha_0+\sigma}\right|\frac{B_{\text{CUT}}^2}{4}\right)\;]\;\}$$

Both these integrals can be further simplified by using the following approximate expressions:

$$j\left[\text{Li}_2(-jx)-\text{Li}_2(jx)\right]\approx$$

$$\approx \pi \,\text{asinh}\left(\frac{x}{2}\right)$$

$$\approx \pi \log_e(1+x)$$

In this case:

$$I_{n_{ch}}^{(n_s)} = -\frac{\pi}{2\alpha_0 B\sigma(2\alpha_0+\sigma)(4\alpha_0+\sigma)}\{(2\alpha_0+\sigma)(-2\alpha_1+\sigma)(4\alpha_0-2\alpha_1+\sigma)$$

$$[\;\text{asinh}\left(\frac{1}{2}\left|\frac{B}{2\alpha_0}\right|\left[f_{n_{ch}}^{(n_s)}-f_{\text{CUT}}-\frac{B_{n_{ch}}^{(n_s)}}{2}\right]\frac{B_{\text{CUT}}}{2}\right)-\text{asinh}\left(\frac{1}{2}\left|\frac{B}{2\alpha_0}\right|\left[f_{n_{ch}}^{(n_s)}-f_{\text{CUT}}+\frac{B_{n_{ch}}^{(n_s)}}{2}\right]\frac{B_{\text{CUT}}}{2}\right)\;]$$

$$+8\alpha_0\alpha_1(2\alpha_0-\alpha_1+\sigma)$$

$$[\;\text{asinh}\left(\frac{1}{2}\left|\frac{B}{2\alpha_0+\sigma}\right|\left[f_{n_{ch}}^{(n_s)}-f_{\text{CUT}}-\frac{B_{n_{ch}}^{(n_s)}}{2}\right]\frac{B_{\text{CUT}}}{2}\right)-\text{asinh}\left(\frac{1}{2}\left|\frac{B}{2\alpha_0+\sigma}\right|\left[f_{n_{ch}}^{(n_s)}-f_{\text{CUT}}+\frac{B_{n_{ch}}^{(n_s)}}{2}\right]\frac{B_{\text{CUT}}}{2}\right)\;]\;\}$$

$$I_{\text{CUT}}^{(n_s)} = \frac{\pi}{\alpha_0 B \sigma (2\alpha_0 + \sigma)(4\alpha_0 + \sigma)}$$

$$\left\{ (2\alpha_0 + \sigma)(-2\alpha_1 + \sigma)(4\alpha_0 - 2\alpha_1 + \sigma) \operatorname{asinh}\left( \frac{1}{2} \left| \frac{B}{2\alpha_0} \right| \frac{B_{\text{CUT}}^2}{4} \right) \right.$$

$$\left. + 8\alpha_0 \alpha_1 (2\alpha_0 - \alpha_1 + \sigma) \operatorname{asinh}\left( \frac{1}{2} \left| \frac{B}{2\alpha_0 + \sigma} \right| \frac{B_{\text{CUT}}^2}{4} \right) \right\}$$

Before re-writing the formulas in full notation, we have one more calculation to accomplish. The total NLI PSD at the end of the link, at the frequency of the CUT, is, according to Eq. 3:

$$G_{\text{NLI}}^{\text{end}}(f_{\text{CUT}}) = \sum_{n_s=1}^{N_s} G_{\text{NLI}}^{(n_s)}(f_{\text{CUT}}) \cdot \left| H(f_{\text{CUT}}; n_s + 1, N_s) \right|^2$$

We still need to evaluate $\left| H(f_{\text{CUT}}; n_s + 1, N_s) \right|^2$. From Eq. 10, it is:

$$\left| H(f_{\text{CUT}}; n_s + 1, N_s) \right|^2 = \prod_{n=n_s+1}^{N_s} \Gamma^{(n)}(f_{\text{CUT}}) e^{-2 \int_0^{L_s^{(n)}} \alpha^{(n)}(f_{\text{CUT}}, z) dz}$$

$$= \prod_{n=n_s+1}^{N_s} \Gamma^{(n)}(f_{\text{CUT}}) e^{-2 \int_0^{L_s^{(n)}} \left[ \alpha_0^{(n)}(f_{\text{CUT}}) + \alpha_1^{(n)}(f_{\text{CUT}}) \exp\left(-\sigma^{(n)}(f_{\text{CUT}}) \cdot z\right) \right] dz}$$

$$= \prod_{n=n_s+1}^{N_s} \Gamma^{(n)}(f_{\text{CUT}}) e^{-2\alpha_0^{(n)}(f_{\text{CUT}}) L_s^{(n)}} e^{2\alpha_1^{(n)}(f_{\text{CUT}}) \left( e^{-\sigma^{(n)}(f_{\text{CUT}}) L_s^{(n)}} - 1 \right) / \sigma^{(n)}(f_{\text{CUT}})}$$

*Eq. 32*

which completes the needed integral calculations. We are now going to re-propose the final formulas, using the full notation.

## 5. Complete formulas in full notation

$$G_{\text{NLI}}^{\text{end}}(f_{\text{CUT}}) = \sum_{n_s=1}^{N_s} G_{\text{NLI}}^{(n_s)}(f_{\text{CUT}}) \cdot \prod_{h=n_s+1}^{N_s} \Gamma^{(h)}(f_{\text{CUT}}) e^{-2\alpha_0^{(h)}(f_{\text{CUT}}) L_s^{(h)}} e^{2\alpha_1^{(h)}(f_{\text{CUT}}) \left( e^{-\sigma^{(h)}(f_{\text{CUT}}) L_s^{(h)}} - 1 \right) / \sigma^{(h)}(f_{\text{CUT}})}$$

*Eq. 33*

$$G_{\text{NLI}}^{(n_s)}(f_{\text{CUT}}) = \frac{16}{27} \left[ \gamma^{(n_s)} \right]^2 \Gamma^{(n_s)}(f_{\text{CUT}}) e^{-2\alpha_0^{(n_s)}(f_{\text{CUT}}) \cdot L_s^{(n_s)}} e^{2\alpha_1^{(n_s)}(f_{\text{CUT}}) \left( e^{-\sigma^{(n_s)}(f_{\text{CUT}}) L_s^{(n_s)}} - 1 \right) / \sigma^{(n_s)}(f_{\text{CUT}})}$$

$$G_{\text{CUT}}^{(n_s)} \left[ \sum_{\substack{n_{\text{ch}}=1 \\ n_{\text{ch}} \neq n_{\text{CUT}}}}^{N_{\text{ch}}^{(n)}} \left[ G_{\text{WDM}, n_{\text{ch}}}^{(n_s)} \right]^2 2 I_{n_{\text{ch}}}^{(n_s)} + \left[ G_{\text{CUT}}^{(n_s)} \right]^2 I_{n_{\text{CUT}}}^{(n_s)} \right]$$

*Eq. 34*

## Integrals with the asinh approximation

$$I_{n_{ch}}^{(n_s)} = -\frac{\pi}{2\alpha_0^{(n_s)}\left(f_{n_{ch}}^{(n_s)}\right)4\pi^2\beta_{2\text{eff},n_{ch}}^{(n_s)}\sigma^{(n_s)}\left(f_{n_{ch}}^{(n_s)}\right)\left(2\alpha_0^{(n_s)}\left(f_{n_{ch}}^{(n_s)}\right)+\sigma^{(n_s)}\left(f_{n_{ch}}^{(n_s)}\right)\right)\left(4\alpha_0^{(n_s)}\left(f_{n_{ch}}^{(n_s)}\right)+\sigma^{(n_s)}\left(f_{n_{ch}}^{(n_s)}\right)\right)}$$

$$\left\{\left(2\alpha_0^{(n_s)}\left(f_{n_{ch}}^{(n_s)}\right)+\sigma^{(n_s)}\left(f_{n_{ch}}^{(n_s)}\right)\right)\left(-2\alpha_1^{(n_s)}\left(f_{n_{ch}}^{(n_s)}\right)+\sigma^{(n_s)}\left(f_{n_{ch}}^{(n_s)}\right)\right)\left(4\alpha_0^{(n_s)}\left(f_{n_{ch}}^{(n_s)}\right)-2\alpha_1^{(n_s)}\left(f_{n_{ch}}^{(n_s)}\right)+\sigma^{(n_s)}\left(f_{n_{ch}}^{(n_s)}\right)\right)\right.$$

$$\left[\text{asinh}\left(\frac{1}{2}\left|\frac{4\pi^2\beta_{2\text{eff},n_{ch}}^{(n_s)}}{2\alpha_0^{(n_s)}\left(f_{n_{ch}}^{(n_s)}\right)}\right|\left[f_{n_{ch}}^{(n_s)}-f_{\text{CUT}}-\frac{B_{n_{ch}}^{(n_s)}}{2}\right]\frac{B_{\text{CUT}}}{2}\right)-\right.$$

$$\text{asinh}\left(\frac{1}{2}\left|\frac{4\pi^2\beta_{2\text{eff},n_{ch}}^{(n_s)}}{2\alpha_0^{(n_s)}\left(f_{n_{ch}}^{(n_s)}\right)}\right|\left[f_{n_{ch}}^{(n_s)}-f_{\text{CUT}}+\frac{B_{n_{ch}}^{(n_s)}}{2}\right]\frac{B_{\text{CUT}}}{2}\right)\right]$$

$$+8\alpha_0^{(n_s)}\left(f_{n_{ch}}^{(n_s)}\right)\alpha_1^{(n_s)}\left(f_{n_{ch}}^{(n_s)}\right)\left(2\alpha_0^{(n_s)}\left(f_{n_{ch}}^{(n_s)}\right)-\alpha_1^{(n_s)}\left(f_{n_{ch}}^{(n_s)}\right)+\sigma^{(n_s)}\left(f_{n_{ch}}^{(n_s)}\right)\right)$$

$$\left[\text{asinh}\left(\frac{1}{2}\left|\frac{4\pi^2\beta_{2\text{eff},n_{ch}}^{(n_s)}}{2\alpha_0^{(n_s)}\left(f_{n_{ch}}^{(n_s)}\right)+\sigma^{(n_s)}\left(f_{n_{ch}}^{(n_s)}\right)}\right|\left[f_{n_{ch}}^{(n_s)}-f_{\text{CUT}}-\frac{B_{n_{ch}}^{(n_s)}}{2}\right]\frac{B_{\text{CUT}}}{2}\right)-\right.$$

$$\left.\text{asinh}\left(\frac{1}{2}\left|\frac{4\pi^2\beta_{2\text{eff},n_{ch}}^{(n_s)}}{2\alpha_0^{(n_s)}\left(f_{n_{ch}}^{(n_s)}\right)+\sigma^{(n_s)}\left(f_{n_{ch}}^{(n_s)}\right)}\right|\left[f_{n_{ch}}^{(n_s)}-f_{\text{CUT}}+\frac{B_{n_{ch}}^{(n_s)}}{2}\right]\frac{B_{\text{CUT}}}{2}\right)\right]\right\}$$

Eq. 35

$$I_{\text{CUT}}^{(n_s)} = \frac{\pi}{\alpha_0^{(n_s)}(f_{\text{CUT}})4\pi^2\beta_{2\text{eff},n_{\text{CUT}}^{(n_s)}}^{(n_s)}\sigma^{(n_s)}(f_{\text{CUT}})\left(2\alpha_0^{(n_s)}(f_{\text{CUT}})+\sigma^{(n_s)}(f_{\text{CUT}})\right)\left(4\alpha_0^{(n_s)}(f_{\text{CUT}})+\sigma^{(n_s)}(f_{\text{CUT}})\right)}$$

$$\left\{\left(2\alpha_0^{(n_s)}(f_{\text{CUT}})+\sigma^{(n_s)}(f_{\text{CUT}})\right)\left(-2\alpha_1^{(n_s)}(f_{\text{CUT}})+\sigma^{(n_s)}(f_{\text{CUT}})\right)\left(4\alpha_0^{(n_s)}(f_{\text{CUT}})-2\alpha_1^{(n_s)}(f_{\text{CUT}})+\sigma^{(n_s)}(f_{\text{CUT}})\right)\right.$$

$$\text{asinh}\left(\frac{1}{2}\left|\frac{4\pi^2\beta_{2\text{eff},n_{\text{CUT}}^{(n_s)}}^{(n_s)}}{2\alpha_0^{(n_s)}(f_{\text{CUT}})}\right|\frac{B_{\text{CUT}}^2}{4}\right)$$

$$+8\alpha_0^{(n_s)}(f_{\text{CUT}})\alpha_1^{(n_s)}(f_{\text{CUT}})\left(2\alpha_0^{(n_s)}(f_{\text{CUT}})-\alpha_1^{(n_s)}(f_{\text{CUT}})+\sigma^{(n_s)}(f_{\text{CUT}})\right)$$

$$\left.\text{asinh}\left(\frac{1}{2}\left|\frac{4\pi^2\beta_{2\text{eff},n_{\text{CUT}}^{(n_s)}}^{(n_s)}}{2\alpha_0^{(n_s)}(f_{\text{CUT}})+\sigma^{(n_s)}(f_{\text{CUT}})}\right|\frac{B_{\text{CUT}}^2}{4}\right)\right\}$$

Eq. 36

$$\beta_{2\text{eff},n_{ch}}^{(n_s)} = \beta_2^{(n_s)} + \pi\beta_3^{(n_s)}\left[f_{n_{ch}}^{(n_s)} + f_{\text{CUT}} - 2f_c^{(n_s)}\right]$$

$$\beta_{2\text{eff},n_{\text{CUT}}^{(n_s)}}^{(n_s)} = \beta_2^{(n_s)} + \pi\beta_3^{(n_s)}\left[2f_{\text{CUT}} - 2f_c^{(n_s)}\right]$$

Eq. 37

## Integrals without the asinh approximation

$$I_{n_{ch}}^{(n_s)} = -\frac{j}{2\alpha_0^{(n_s)}\left(f_{n_{ch}}^{(n_s)}\right)4\pi^2\beta_{2\text{eff},n_{ch}}^{(n_s)}\sigma^{(n_s)}\left(f_{n_{ch}}^{(n_s)}\right)\left(2\alpha_0^{(n_s)}\left(f_{n_{ch}}^{(n_s)}\right)+\sigma^{(n_s)}\left(f_{n_{ch}}^{(n_s)}\right)\right)\left(4\alpha_0^{(n_s)}\left(f_{n_{ch}}^{(n_s)}\right)+\sigma^{(n_s)}\left(f_{n_{ch}}^{(n_s)}\right)\right)}$$

$$\left\{\left(2\alpha_0^{(n_s)}\left(f_{n_{ch}}^{(n_s)}\right)+\sigma^{(n_s)}\left(f_{n_{ch}}^{(n_s)}\right)\right)\left(-2\alpha_1^{(n_s)}\left(f_{n_{ch}}^{(n_s)}\right)+\sigma^{(n_s)}\left(f_{n_{ch}}^{(n_s)}\right)\right)\left(4\alpha_0^{(n_s)}\left(f_{n_{ch}}^{(n_s)}\right)-2\alpha_1^{(n_s)}\left(f_{n_{ch}}^{(n_s)}\right)+\sigma^{(n_s)}\left(f_{n_{ch}}^{(n_s)}\right)\right)\right.$$

$$\left[\text{Li}_2\left(-j\left|\frac{4\pi^2\beta_{2\text{eff},n_{ch}}^{(n_s)}}{2\alpha_0^{(n_s)}\left(f_{n_{ch}}^{(n_s)}\right)}\right|\left[f_{n_{ch}}^{(n_s)}-f_{\text{CUT}}-\frac{B_{n_{ch}}^{(n_s)}}{2}\right]\frac{B_{\text{CUT}}}{2}\right)-\text{Li}_2\left(j\left|\frac{4\pi^2\beta_{2\text{eff},n_{ch}}^{(n_s)}}{2\alpha_0^{(n_s)}\left(f_{n_{ch}}^{(n_s)}\right)}\right|\left[f_{n_{ch}}^{(n_s)}-f_{\text{CUT}}-\frac{B_{n_{ch}}^{(n_s)}}{2}\right]\frac{B_{\text{CUT}}}{2}\right)\right.$$

$$\left.-\text{Li}_2\left(-j\left|\frac{4\pi^2\beta_{2\text{eff},n_{ch}}^{(n_s)}}{2\alpha_0^{(n_s)}\left(f_{n_{ch}}^{(n_s)}\right)}\right|\left[f_{n_{ch}}^{(n_s)}-f_{\text{CUT}}+\frac{B_{n_{ch}}^{(n_s)}}{2}\right]\frac{B_{\text{CUT}}}{2}\right)+\text{Li}_2\left(j\left|\frac{4\pi^2\beta_{2\text{eff},n_{ch}}^{(n_s)}}{2\alpha_0^{(n_s)}\left(f_{n_{ch}}^{(n_s)}\right)}\right|\left[f_{n_{ch}}^{(n_s)}-f_{\text{CUT}}+\frac{B_{n_{ch}}^{(n_s)}}{2}\right]\frac{B_{\text{CUT}}}{2}\right)\right]$$

$$+8\alpha_0^{(n_s)}\left(f_{n_{ch}}^{(n_s)}\right)\alpha_1^{(n_s)}\left(f_{n_{ch}}^{(n_s)}\right)\left(2\alpha_0^{(n_s)}\left(f_{n_{ch}}^{(n_s)}\right)-\alpha_1^{(n_s)}\left(f_{n_{ch}}^{(n_s)}\right)+\sigma^{(n_s)}\left(f_{n_{ch}}^{(n_s)}\right)\right)$$

$$\left[\text{Li}_2\left(-j\left|\frac{4\pi^2\beta_{2\text{eff},n_{ch}}^{(n_s)}}{2\alpha_0^{(n_s)}\left(f_{n_{ch}}^{(n_s)}\right)+\sigma^{(n_s)}\left(f_{n_{ch}}^{(n_s)}\right)}\right|\left[f_{n_{ch}}^{(n_s)}-f_{\text{CUT}}-\frac{B_{n_{ch}}^{(n_s)}}{2}\right]\frac{B_{\text{CUT}}}{2}\right)-\right.$$

$$\text{Li}_2\left(j\left|\frac{4\pi^2\beta_{2\text{eff},n_{ch}}^{(n_s)}}{2\alpha_0^{(n_s)}\left(f_{n_{ch}}^{(n_s)}\right)+\sigma^{(n_s)}\left(f_{n_{ch}}^{(n_s)}\right)}\right|\left[f_{n_{ch}}^{(n_s)}-f_{\text{CUT}}-\frac{B_{n_{ch}}^{(n_s)}}{2}\right]\frac{B_{\text{CUT}}}{2}\right)$$

$$-\text{Li}_2\left(-j\left|\frac{4\pi^2\beta_{2\text{eff},n_{ch}}^{(n_s)}}{2\alpha_0^{(n_s)}\left(f_{n_{ch}}^{(n_s)}\right)+\sigma^{(n_s)}\left(f_{n_{ch}}^{(n_s)}\right)}\right|\left[f_{n_{ch}}^{(n_s)}-f_{\text{CUT}}+\frac{B_{n_{ch}}^{(n_s)}}{2}\right]\frac{B_{\text{CUT}}}{2}\right)+$$

$$\left.\left.\text{Li}_2\left(j\left|\frac{4\pi^2\beta_{2\text{eff},n_{ch}}^{(n_s)}}{2\alpha_0^{(n_s)}\left(f_{n_{ch}}^{(n_s)}\right)+\sigma^{(n_s)}\left(f_{n_{ch}}^{(n_s)}\right)}\right|\left[f_{n_{ch}}^{(n_s)}-f_{\text{CUT}}+\frac{B_{n_{ch}}^{(n_s)}}{2}\right]\frac{B_{\text{CUT}}}{2}\right)\right]\right\}$$

*Eq. 38*

$$I_{\text{CUT}}^{(n_s)} = \frac{j}{\alpha_0^{(n_s)}\left(f_{\text{CUT}}\right)4\pi^2\beta_{2\text{eff},n_{\text{CUT}}^{(n_s)}}^{(n_s)}\sigma^{(n_s)}\left(f_{\text{CUT}}\right)\left(2\alpha_0^{(n_s)}\left(f_{\text{CUT}}\right)+\sigma^{(n_s)}\left(f_{\text{CUT}}\right)\right)\left(4\alpha_0^{(n_s)}\left(f_{\text{CUT}}\right)+\sigma^{(n_s)}\left(f_{\text{CUT}}\right)\right)}$$

$$\left\{\left(2\alpha_0^{(n_s)}\left(f_{\text{CUT}}\right)+\sigma^{(n_s)}\left(f_{\text{CUT}}\right)\right)\left(-2\alpha_1^{(n_s)}\left(f_{\text{CUT}}\right)+\sigma^{(n_s)}\left(f_{\text{CUT}}\right)\right)\left(4\alpha_0^{(n_s)}\left(f_{\text{CUT}}\right)-2\alpha_1^{(n_s)}\left(f_{\text{CUT}}\right)+\sigma^{(n_s)}\left(f_{\text{CUT}}\right)\right)\right.$$

$$\left[\text{Li}_2\left(-j\left|\frac{4\pi^2\beta_{2\text{eff},n_{\text{CUT}}^{(n_s)}}^{(n_s)}}{2\alpha_0^{(n_s)}\left(f_{\text{CUT}}\right)}\right|\frac{B_{\text{CUT}}^2}{4}\right)-\text{Li}_2\left(j\left|\frac{4\pi^2\beta_{2\text{eff},n_{\text{CUT}}^{(n_s)}}^{(n_s)}}{2\alpha_0^{(n_s)}\left(f_{\text{CUT}}\right)}\right|\frac{B_{\text{CUT}}^2}{4}\right)\right]$$

$$+8\alpha_0\alpha_1\left(2\alpha_0-\alpha_1+\sigma\right)$$

$$\left.\left[\text{Li}_2\left(-j\left|\frac{4\pi^2\beta_{2\text{eff},n_{\text{CUT}}^{(n_s)}}^{(n_s)}}{2\alpha_0^{(n_s)}\left(f_{\text{CUT}}\right)+\sigma^{(n_s)}\left(f_{\text{CUT}}\right)}\right|\frac{B_{\text{CUT}}^2}{4}\right)-\text{Li}_2\left(j\left|\frac{4\pi^2\beta_{2\text{eff},n_{\text{CUT}}^{(n_s)}}^{(n_s)}}{2\alpha_0^{(n_s)}\left(f_{\text{CUT}}\right)+\sigma^{(n_s)}\left(f_{\text{CUT}}\right)}\right|\frac{B_{\text{CUT}}^2}{4}\right)\right]\right\}$$

*Eq. 39*

$$\beta^{(n_s)}_{2\text{eff},n_{\text{ch}}} = \beta^{(n_s)}_2 + \pi\beta^{(n_s)}_3\left[f^{(n_s)}_{n_{\text{ch}}} + f_{\text{CUT}} - 2f^{(n_s)}_c\right]$$

$$\beta^{(n_s)}_{2\text{eff},n^{(n_s)}_{\text{CUT}}} = \beta^{(n_s)}_2 + \pi\beta^{(n_s)}_3\left[2f_{\text{CUT}} - 2f^{(n_s)}_c\right]$$

## 6. Special values of certain parameters

It may occur that $\beta^{(n_s)}_{2\text{eff},n_{\text{ch}}} = 0$ for some index $n_{\text{ch}}$, including for the CUT, that is $\beta^{(n_s)}_{2\text{eff},n^{(n_s)}_{\text{CUT}}} = 0$. If so, the results above become a 0/0 form. To solve for these special cases, we go back to the integrals and find the result directly:

$$I^{(n_s)}_{n_{\text{ch}}}\bigg|_{\beta^{(n_s)}_{2,n_{\text{ch}}}=0} \approx \frac{B^{(n_s)}_{n_{\text{CUT}}} B^{(n_s)}_{n_{\text{ch}}}}{4\left[\alpha^{(n_s)}_0\left(f^{(n_s)}_{n_{\text{ch}}}\right)\right]^2} \cdot \frac{\left(2\alpha^{(n_s)}_0\left(f^{(n_s)}_{n_{\text{ch}}}\right) - 2\alpha^{(n_s)}_1\left(f^{(n_s)}_{n_{\text{ch}}}\right) + \sigma^{(n_s)}\left(f^{(n_s)}_{n_{\text{ch}}}\right)\right)^2}{\left(2\alpha^{(n_s)}_0\left(f^{(n_s)}_{n_{\text{ch}}}\right) + \sigma^{(n_s)}\left(f^{(n_s)}_{n_{\text{ch}}}\right)\right)^2}$$

$$I^{(n_s)}_{\text{CUT}} \approx \frac{\left[B^{(n_s)}_{n_{\text{CUT}}}\right]^2}{4\left[\alpha^{(n_s)}_0\left(f_{\text{CUT}}\right)\right]^2} \cdot \frac{\left(2\alpha^{(n_s)}_0\left(f_{\text{CUT}}\right) - 2\alpha^{(n_s)}_1\left(f_{\text{CUT}}\right) + \sigma^{(n_s)}\left(f_{\text{CUT}}\right)\right)^2}{\left(2\alpha^{(n_s)}_0\left(f_{\text{CUT}}\right) + \sigma^{(n_s)}\left(f_{\text{CUT}}\right)\right)^2}$$

Note that these results are still derived assuming a first-order series expansion in $\alpha_1$ and hence they too assume $\alpha_1 \ll \alpha_0$.

## 7. Setting the SRS parameters

As shown above, loss for each channel is modeled as:

$$\alpha^{(n_s)}\left(f^{(n_s)}_{n_{\text{ch}}}, z\right) = \alpha^{(n_s)}_0\left(f^{(n_s)}_{n_{\text{ch}}}\right) + \alpha^{(n_s)}_1\left(f^{(n_s)}_{n_{\text{ch}}}\right)\exp\left(-\sigma^{(n_s)}\left(f^{(n_s)}_{n_{\text{ch}}}\right)\cdot z\right)$$

The fiber loss is easily obtained through direct measurements or from the literature. Regarding $\alpha^{(n_s)}_1\left(f^{(n_s)}_{n_{\text{ch}}}\right)$ and $\sigma^{(n_s)}\left(f^{(n_s)}_{n_{\text{ch}}}\right)$ they can be obtained based on the existing SRS theory. A number of results is readily available, dating back for instance to 1998 [14]. Depending on the level of accuracy and complexity, relatively simple parametrization can be carried out. For instance, the values of $\alpha^{(n_s)}_1\left(f^{(n_s)}_{n_{\text{ch}}}\right)$, which correspond the extra loss/gain due to SRS at the start of the span, can be easily calculated from [14]. Regarding the decay constant $\sigma^{(n_s)}\left(f^{(n_s)}_{n_{\text{ch}}}\right)$, simple approximations (applying the same value to all channels based on some average value) or more sophisticated calculations based on the actual spectral distribution of power can be used.

## 8. Conclusion

In this paper we have outlined the full derivation of an NLI formula based on the incoherent GN-model (the iGN-model) which extends the previously available result from [4], within the same analytical framework, to encompass frequency-dependent dispersion, frequency-dependent loss and SRS. This result allows to carry out more accurate evaluation of network performance in real-time, even in complex scenarios, allowing effective on-the-fly control, management and optimization actions.

## 9. Acknowledgements

This work was sponsored by CISCO Photonics under an SRA agreement with Politecnico di Torino. We author would like to thank Stefano Piciaccia and Fabrizio Forghieri from CISCO Photonics for the fruitful discussions and interactions.